\documentclass[fleqn,usenatbib]{mnras}
\usepackage{newtxtext,newtxmath}
\usepackage[normalem]{ulem}
\usepackage[T1]{fontenc}
\usepackage{ae,aecompl}

\usepackage{graphicx}	
\usepackage{amssymb}	
\usepackage{xcolor}
\usepackage{hyperref}
\hypersetup{colorlinks=True, linkcolor=blue!50!black, citecolor=black,
  urlcolor=blue!50!black}
\usepackage{hyphenat}
\def\apj{ApJ}
\def\aj{AJ}
\def\mnras{MNRAS}

\def\aap{A\&A}
\def\apjs{ApJS}

\newcommand{\kms}{\ensuremath{\mathrm{km\ s}^{-1}}}

\newcommand{\nii}{[\ion{N}{II}]}

\newcommand{\oiii}{[\ion{O}{III}]}

\newcommand\NIIlam{\nii\ \(\lambda 6584\)}
\newcommand\NIIlamlam{\nii\ \(\lambda\lambda6548,84\)}
\newcommand\OIII{[\ion{O}{III}]}
\newcommand\OIIIlam{\oiii\ \(\lambda 5007\)}
\newcommand\OIIIlamlam{\oiii\ \(\lambda\lambda 4959,5007\)}

\newcommand\Halam{\Ha\ \(\lambda 6563\)}
\newcommand\Hblam{\Hb\ \(\lambda 4861\)}
\newcommand\Ha{\ensuremath{\mathrm{H}\alpha}}
\newcommand\Hb{\ensuremath{\mathrm{H}\beta}}
\newcommand{\vhel}{\ensuremath{V_\mathrm{hel}}}

\title[The Owl and other strigiform nebulae]{The Owl and other strigiform nebulae:\\ multipolar cavities within a filled shell\thanks{%
Based on observations carried out at the Observatorio Astron\'omico
Nacional, Sierra San Pedro M\'artir (OAN-SPM), Baja
California, Mexico, operated by the Universidad Nacional
Aut\'onoma de M\'exico.}}

\author[Garc\'{\i}a-D\'{\i}az et al.]{%
Ma.~T. Garc\'{\i}a-D\'{\i}az,$^{1}$\thanks{%
E-mail: tere@astro.unam.mx},
W. Steffen,$^{1}$
W. J. Henney,$^{2}$ J. A. L\'opez,$^{1}$
F.~Garc\'ia-L\'opez,$^{3}$
\newauthor    
D.~Gonz\'alez-Buitrago$^{4}$ 
and A. \'Aviles$^{5}$\\
  $^{1}$ Instituto de Astronom\'ia, Universidad Na\-cio\-nal
    Aut\'o\-no\-ma de M\'e\-xico. Km 103 Carretera Tijuana-Ensenada,
    22860 Ensenada,\\ Baja California, M\'exico\\
 $^{2}$ Instituto de Radioastronom\'ia y Astrof\'isica,
   Universidad Nacional Aut\'o\-noma de M\'e\-xico, Apartado Postal
    3-72, 58090 Morelia,\\ Michoac\'an, M\'exico\\
 $^{3}$  Centro de Nanociencias y Nanotecnolog\'ia, UNAM\\
  $^{4}$ Department of Physics and Astronomy, 4129 Frederick Reines Hall, University of California, Irvine, CA 92697, USA\\
  $^{5}$ Facultad de Ingenier\'ia Mec\'anica y El\'ectrica
   de la Universidad Aut\'onoma de Nuevo Le\'on
   }

\date{Accepted XXX. Received YYY; in original form ZZZ}

\pubyear{2018}

\begin{document}

\date{Submitted: \today}

\pagerange{\pageref{firstpage}--\pageref{lastpage}} \pubyear{2018}

\maketitle

\label{firstpage}

\begin{abstract}
We present the results of long-slit echelle spectroscopy and deep narrow-band imaging of the Owl Nebula (NGC~3587), obtained at the \textit{Observatorio Astron\'omico Nacional, San Pedro M\'artir}. 
These data allow us to construct an iso-velocity data cube and develop a 3-D morpho-kinematic model.  
We find that, instead of the previously assumed bipolar dumbbell shape, the inner cavity consists of multi-polar fingers within an overall tripolar structure.  
We identify three additional planetary nebulae that show very similar morphologies and kinematics to the Owl, and propose that these constitute a new class of \textit{strigiform} (owl-like) nebulae. 
Common characteristics of the strigiform nebulae include a double-shell (thin outside thick) structure, low-luminosity and high-gravity central stars, the absence of a present-day stellar wind, and asymmetric inner cavities, visible in both optical and mid-infrared emission lines, that show no evidence for surrounding bright rims.
The origin of the cavities is unclear, but they may constitute relics of an earlier stage of evolution when the stellar wind was active.
\end{abstract}


\section{Introduction}
\label{sec:introduction}
NGC~3587 is better known as the Owl Nebula, so named because it shows two holes in its bright emission disk that resemble the eyes of an owl.  These eyes constitute a seemingly bipolar internal cavity within a roughly spherical overall structure.\footnote{%
However, we will show in this paper that such an interpretation is not strictly correct and that the true 3D geometry is tripolar or multipolar.}  
The nebula has been studied using deep narrow-band imaging and low resolution spectroscopy with good spatial coverage (e.g., \citealp{Chu:1987}; \citealp{Kwitter:1993a}; \citealp{Cuesta:2000a}). 
The nebula has also been studied using high-resolution longslit spectroscopy  in the lines of \Ha, \OIIIlam\, and \NIIlam{} (\citealp{Sabbadin:1985a}; \citealp{Guerrero:2003a}), but only at a limited number of position angles. 
No diffuse x-rays have been detected from the nebula \citep{Kastner:2012a}, but this is unsurprising for such a large, evolved object.  
The faint outer halo of the nebula has been studied via integral field spectroscopy \citep{Sandin:2008a}. 

The central star of NGC 3587 has a spectroscopically determined effective temperature \(T_{\mathrm{eff}} = (94 \pm 6)~\mathrm{kK}\) and gravity \(\log g = 6.9 \pm 0.3\) \citep{Napiwotzki:1999a}, which implies a luminosity \mbox{\(L = 140 \pm 100 \ \mathrm{L_\odot}\)}.  
Similar values are found independently via the Zanstra method \citep{Gorny:1997a}, which yields \(T_{\mathrm{eff}} \approx 110~\mathrm{kK}\) and \(L \approx 70~\mathrm{L_\odot}\).  
These \((L, T_{\mathrm{eff}})\) values are consistent with post-AGB evolutionary tracks \citep{Miller-Bertolami:2016a} for a range of remnant stellar masses less than \(M = 0.65~\mathrm{M_\odot}\), but the predicted evolutionary ages cover a very wide range, from \(t = 5000\) to \(10^5~\mathrm{yr}\), with lower masses and higher metallicities corresponding to larger ages.   
X-ray emission from the central star has been detected by ROSAT \citep{Chu:1998a}, with a spectrum consistent with \(T_{\mathrm{eff}} \sim 100~\mathrm{kK}\) photospheric emission, but it was not detected by Chandra \citep{Kastner:2012a}.

The most reliable distance estimates for the Owl are given in \citet{Frew:2016a}.  
From the stellar parameters described above they find a  ``gravity distance" of \(0.87 \pm 0.26\)~kpc, while from the nebular H\(\alpha\) flux they find a statistical distance via the surface brightness--radius relation of \(0.79 \pm 0.22\)~kpc. 

The basic physical parameters of the nebula are well established from the above-mentioned studies.  
The main body of the nebula comprises an inner shell, of radius \(\approx 70''\) (0.3~pc), most prominent in high-ionization lines, and a fainter outer shell, of radius \(\approx 100''\) (0.44~pc), most prominent in low-ionization lines.  
Both shells are approximately circular in appearance, with the outer shell having a knotty appearance and the inner shell being smoother, apart from the low-intensity cavity structures that constitute the ``eyes" of the Owl and which define the major axis at PA ($\approx -45 \degr$).   
There is also a much fainter outer halo \citep{Kwitter:1993a, Sandin:2008a}, which extends out to radii of \(180''\) (0.8~pc) and which shows evidence for a bowshock-type interaction with the interstellar medium \citep{Guerrero:2003a}.  
The nebula is located at a high Galactic latitude and the foreground dust extinction is negligible \citep{Kaler:1976a}.

\newcounter{IONcounter}
\newcommand\ION[2]{\setcounter{IONcounter}{#2}#1\,\textsc{\roman{IONcounter}}}

Electron temperatures have been measured  from collisionally excited line ratios at optical wavelengths \citep{Boeshaar:1974a, Torres-Peimbert:1977a, Kwitter:2001a, Sandin:2008a}, finding 10,500--11,500~K, both for high-ionization ([\ion{O}{III}], [\ion{S}{III}]) and low-ionization  ([\ion{N}{II}], [\ion{O}{II}]) zones.     
Electron densities have been measured for the low-ionization zone via density-sensitive doublet ratios of [\ion{S}{II}] and [\ion{O}{II}] \citep{Osterbrock:1960a, Boeshaar:1974a, Stanghellini:1989a, Sandin:2008a}, finding values of 50--150~\(\mathrm{cm}^{-3}\).  
The results have a large uncertainty because the observed line ratios are close to the low-density limit of 1.5, indeed NGC~3587 has one of the lowest densities of any observed planetary nebula \citep{Wang:2004a}.\footnote{%
	An outlier is the study of \citet{Cuesta:2000a}, who claim to measure a mean [\ion{S}{II}] electron density of \(\approx 600~\mathrm{cm}^{-3}\), with strong gradients across the nebula.  As noted by \citet{Sandin:2008a}, this is completely inconsistent with all other observations of the nebula.   However, the [\ion{S}{II}] intensity profiles shown in \citeauthor{Cuesta:2000a} Fig.~4 imply 6717/6731 ratios of 1.3--1.5, which \emph{are} consistent with other studies and imply densities  \(< 100~\mathrm{cm}^{-3}\).  This suggests that the authors mistake lay in the derivation of density from the line ratios.}  
Unfortunately, no [\ion{Cl}{III}] or [\ion{Ar}{IV}] doublet ratios have been published that would allow densities in the high-ionization zone to be determined. 

An alternative method for estimating the electron density is to derive the emission measure from the  H\(\alpha\) surface brightness and the known temperature-dependent line emission coefficient \citep{Osterbrock:2006a}, from which the electron density follows for an assumed line-of-sight emission depth and filling factor.   
Taking the average surface brightness of the inner shell, \((1.4 \pm 0.2) \times 10^{-4}~\mathrm{erg\ s^{-1}\ cm^{-2}\ sr^{-1}}\) \citep{Frew:2016a}, and assuming an emission depth equal to the inner-shell diameter of 0.6~pc and filling factor of \(\varepsilon\) yields a density of  \((50 \pm 10) \,\varepsilon^{-1/2}~\mathrm{cm^{-3}}\), which is fully consistent with the line ratio measurements.  
Applying the same technique to other components of the nebula gives \(\approx 30\,\varepsilon^{-1/2}~\mathrm{cm^{-3}}\) for the outer shell and \(\approx 6\,\varepsilon^{-1/2}~\mathrm{cm^{-3}}\) for the halo. 

The kinematics of the inner shell are relatively simple, with long-slit spectra \citep{Sabbadin:1985a, Guerrero:2003a} showing knotty velocity ellipses with peak-to-peak splitting that is higher for [\ion{N}{II}] (\(80 \pm 2~\mathrm{km\ s^{-1}}\)) than for [\ion{O}{III}] (\(56 \pm 3~\mathrm{km\ s^{-1}}\)), which is consistent with approximately spherical homologous expansion (velocity proportional to radius).  
There is  a velocity asymmetry in the inner-shell emission cavities along the major axis,  which led \citet{Guerrero:2003a} to propose a bipolar model for the cavities.  
There is no sign of limb brightening at the cavity edges.  
The outer shell is kinematically similar, showing clear separation from the inner shell along the minor axis, but merging into it along the major axis \citep{Guerrero:2003a}. 
The halo, in contrast, shows very narrow, unsplit lines, but with a radial gradient in mean velocity.  

In the remainder of this paper we first, in \S~2, present a complete spectro-kinematic study of the Owl Nebula using 12 parallel slit positions in [\ion{N}{II}] and H\(\alpha\).  
This allows a more thorough analysis of the deviations from spherical symmetry than was possible in previous studies, which were restricted to position angles along the major and minor axes.  
and we construct a continuous cube of iso-velocity maps that help in the interpretation of the nebular structure, followed, in \S~3, by a morpho$–$kinematical model using the code \textsc{Shape}, which reproduces the new data in detail.  
Finally, in \S~4, we discuss the interpretation of our results in terms of the evolution of the Owl Nebula and other planetaries.

\begin{figure}
\includegraphics[width=\linewidth]{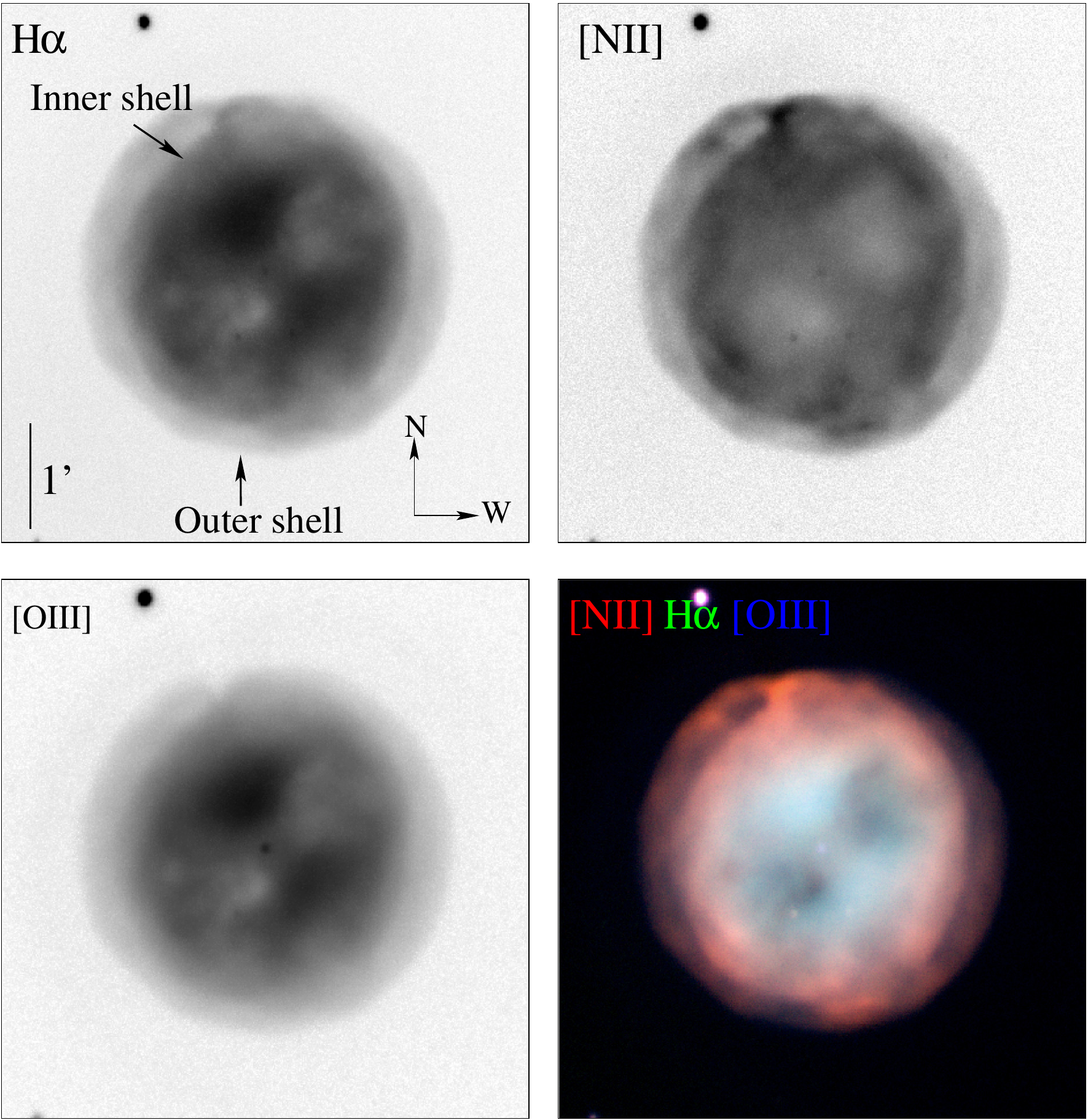}
\caption{Images of NGC 3587 obtained at the OAN-SPM 84~cm
     telescope, in the light of (from upper left)  H$\alpha$, 
     \NIIlam{}\AA{},  and \OIIIlam{}, with lower right panel showing
     RGB composite of the three filters.}
  \label{fig:imagenes}
\end{figure}

\begin{figure}
    \centering
  \includegraphics[width=0.45\textwidth]{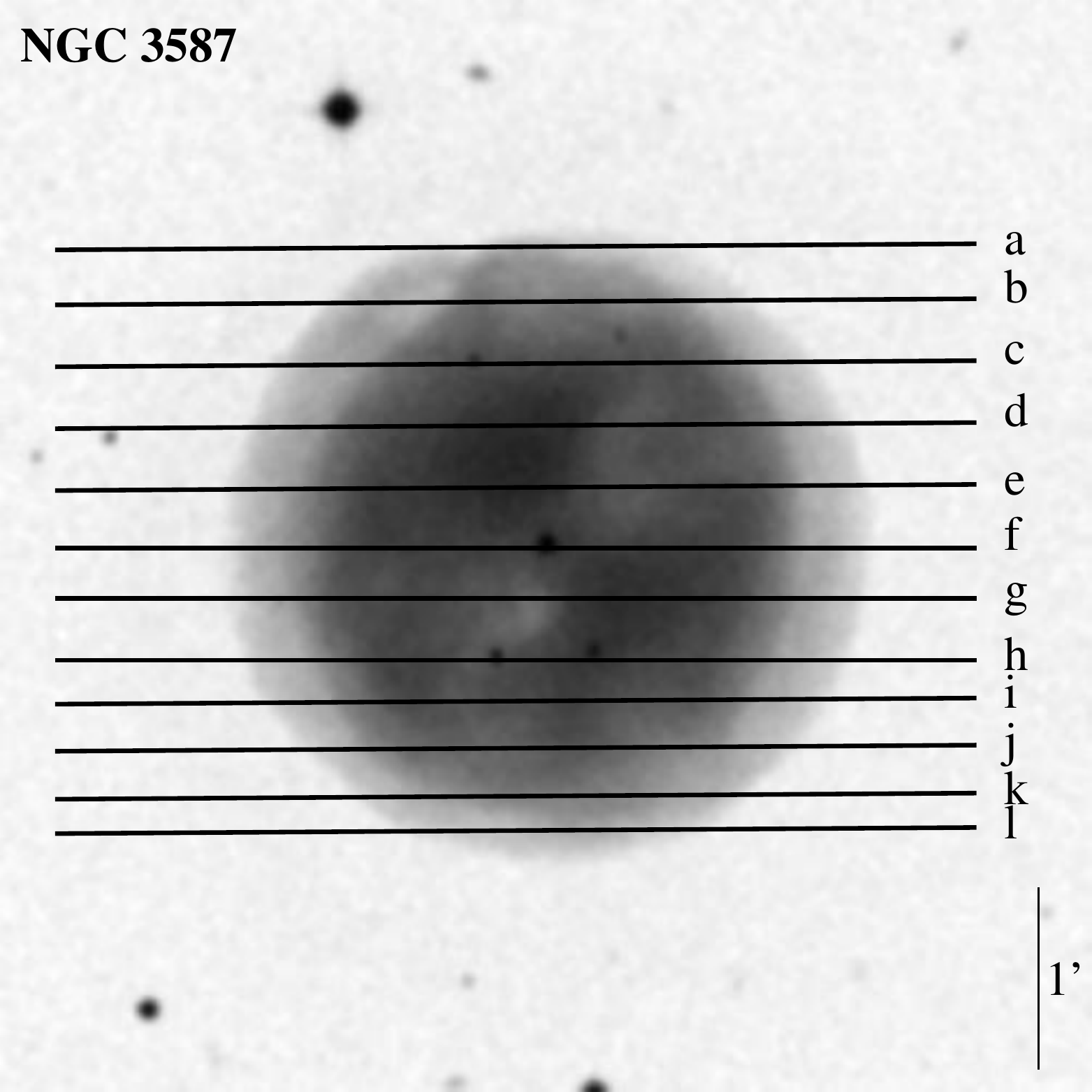}
   \caption{Slit positions for the longslit spectra,  is indicated and labeled on
    an \Ha{} image of NGC 3587. North is up, east
    left.}
  \label{fig:image+slit}
\end{figure}

\begin{table*}
\centering
\caption{Detector characteristics for longslit observations}
\smallskip

\begin{tabular}{lccccccc}\hline
Epoch & Number &slit & CCD & CCD size & Pixel size  & Binning  & Slit length   \\
& of spectra & & & (pixels) & ($\mu$m) & (pixels) & (arcmin) \\  \hline 
2001  &  2 &  f, h &SiTe3 &  1024$\times$1024 & 24.0  & 2$\times$2 & 5.32\\
2013  & 10 &  a,b,c,d,e,g,i,j,k,l& Marconi & 2048$\times$2048 &  13.6  & 3$\times$3 & 5.47\\
\hline
\hline
\end{tabular}
\label{tab:ccds}
\end{table*}

\section{Observations and results}
\label{sec:observations}

The observations of the planetary nebula NGC~3587 were obtained at the
\emph{Observatorio Astron\'omico Nacional} at \emph{San Pedro M\'artir}, (OAN-SPM), Baja California, Mexico, and standard data reduction was performed using IRAF.\footnote{
IRAF is distributed by the National Optical Astronomy Observatories, which is operated by the Association of Universities for Research in Astronomy, Inc. under cooperative agreement with the National Science foundation} 
The reduction steps included  bias correction, cosmic ray removal, and flat-fielding.  
In addition, the long-slit high resolution spectra were wavelength-calibrated by means of comparison lamp spectra.

\subsection{Optical Imaging}
\label{sec:imaging}
A set of narrow-band CCD direct images of NGC 3587 were obtained on
2015 June 22 using the 0.84 m Telescope. The detector was a 2048 $\times$
2048 Marconi CCD (13.5 $\mu$m pixel size). The filters used to acquire
the images were: H$\alpha$ (bandwidth, $\Delta\lambda$ = 11\AA, central
wavelength, $\lambda_c$ = 6565\AA); \OIII\, ($\Delta\lambda$ = 52\AA,
$\lambda_c$ = 5009\AA), and \nii\,
($\Delta\lambda$ = 10\AA, $\lambda_c$ 6585\AA). Exposure times were
1800~s for each filter. The images were aligned with respect to
each other. 

Figure~\ref{fig:imagenes} is a mosaic of ground-based images of NGC~3587, in which the
panels correspond to \Ha{} (top left), \NIIlam{} (top right), \OIIIlam{}
(bottom left), and an RGB color composite image of the nebula (bottom right).  
In the color composite, the inner core of the nebula can be seen to be dominated by \Ha{} and \oiii{} (green and blue, respectively), while both the outer rim of the inner shell (diameter 139\arcsec) and the outer shell (diameter 204\arcsec) are dominated by \nii{} (red).
The \Ha{}  and \oiii{} images show two inner regions on either side of the star where the surface brightness decreases, known as the eyes of the Owl. The spatial resolution of our images is limited to \(\approx 1.5\arcsec\) by seeing conditions and telescope tracking accuracy.  Finer details in the structure of the nebular cavities can be seen in a longer exposure and highly processed image from the Calar Alto Observatory.\footnote{%
\url{http://www.starshadows.com/gallery/display.cfm?imgID=304} Note that north is right and east is the top in that image, i.e. it is rotated 90$\degr$ clockwise from the orientations used in this paper.\label{fn:calar-alto} 
} 

\begin{figure*}
\centering
 \includegraphics[width=1\textwidth]{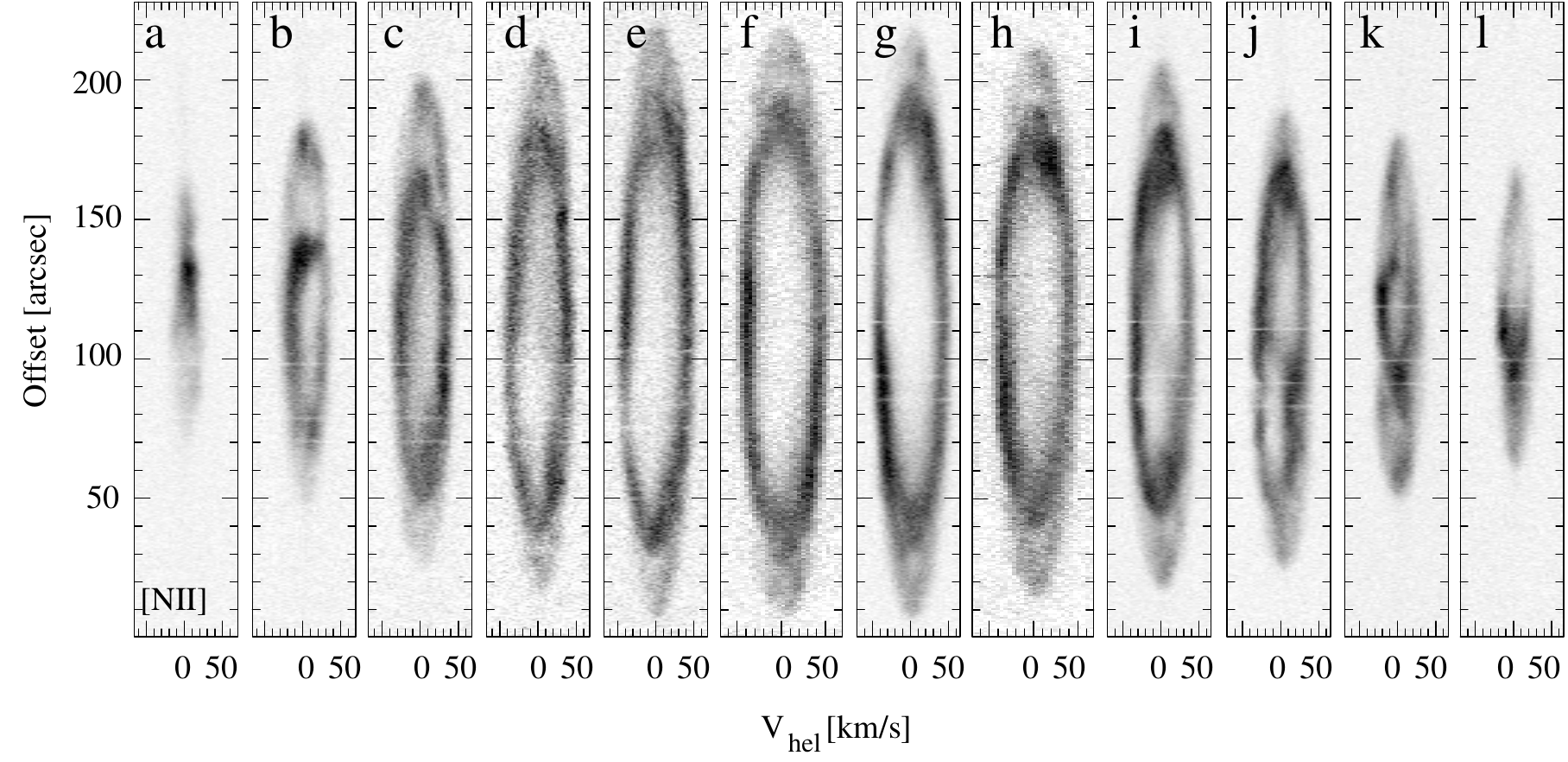}
  \caption{Mosaic of observed \NIIlam{} bi-dimensional P-V arrays labeled according to slit position from  North to South, \textit{a}--\textit{l}.  The top of each P-V array corresponds to the East side of the nebula (see Figure 2).}
    \label{fig:pvobsnii}
\end{figure*}

\begin{figure*}
\centering
 \includegraphics[width=1\textwidth]{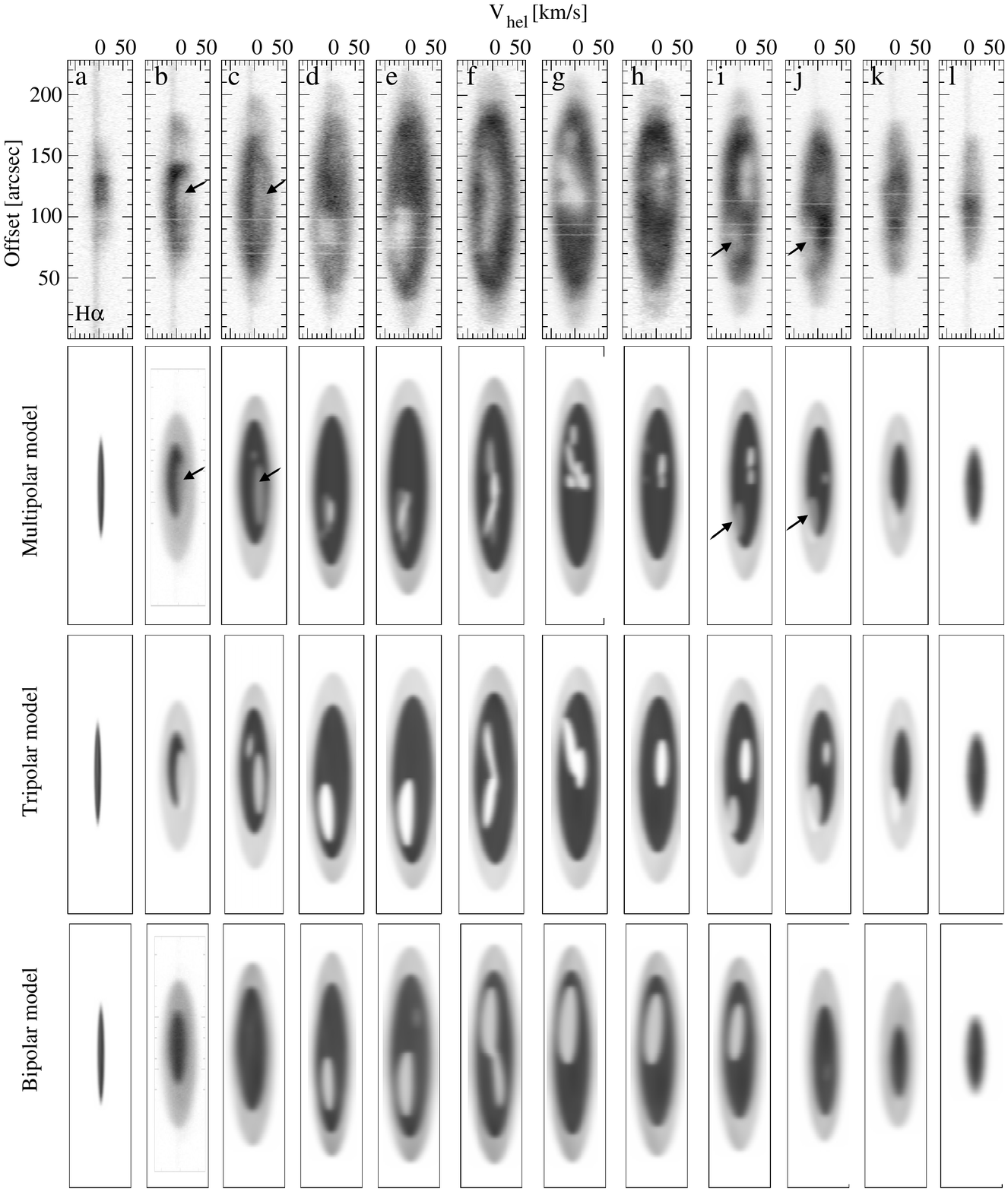}
  \caption{{\it First row:} Mosaic of  bi-dimensional \Ha\, P-V arrays labeled
    according to slit position. {
\it Second row:} Synthetic P-V array derived from a Multipolar model. {\it Third row:} Synthetic P-V array derived from  a tripolar model. {\it Fourth row:} Synthetic P-V array derived from a bipolar model. The top of each P-V array corresponds to the East side of the nebula}
    \label{fig:pvobs}
\end{figure*}

\subsection{High Resolution Spectroscopy}
\label{sec:spectroscopy}

We obtained high-resolution spectroscopic data of the Owl Nebula
during the nights of 2001 May 22 and 2013 February 17--18, employing
the Manchester Echelle Spectrometer (MES-SPM) (\citealt{Meaburn:2003}) on the
2.1~m telescope in a $f$/7.5 configuration. We used a \(\Delta\lambda = 90\)~\AA{} filter to isolate the 87th order, containing the \Halam{} and \NIIlamlam{} nebular emission lines. 
All observations used a 150~$\mu$m wide slit (\(1.9\arcsec\)) which was
oriented east-west and stepped over the face of the Owl nebula from north to south. 
The detector characteristics were different for the two runs and are listed in Table~\ref{tab:ccds}.  
After on-chip binning, the resultant spectra for the 2001 (2013) observations have 512 (682) pixels along each dimension.  
Each binned pixel has a projected angular size \(0.624\arcsec\) (\(0.531\arcsec\)) along the spatial dimension and a velocity width of \(4.6\)~\kms{} (\(4.0\)~\kms{}) along the spectral dimension.  
These give a close to optimum sampling of both the seeing width (FWHM of 1--1.5\arcsec) and the spectrograph profile (FWHM of 12~\kms). 
  
Spectra were obtained at 12 different positions across the nebula, as illustrated in Figure~\ref{fig:image+slit}, with exposure times of 1800~s each, the separation between exposures was 20\arcsec. 
For each pointing, an additional exposure was obtained with the ``image plus slit'' mode of the MES. 
This allowed the astrometric determination of the exact slit position by means of comparison with the positions of nearby stars taken from Digital Sky Survey images.
Wavelength calibration against the spectrum of a Th/Ar lamp was obtained after each science exposure, yielding an accuracy of \(\pm 1\)~\kms{} in the absolute velocities, which were transformed to the heliocentric frame (\vhel). 
The resultant spectra are presented in Figure~\ref{fig:pvobsnii} (\NIIlam) and the top row of Figure~\ref{fig:pvobs} (\Halam). 
Some dust spots on the spectrograph slit give rise to instrumental artifacts that can be seen as narrow horizontal stripes in the spectra shown in the figures. 

Our kinematic data add crucial new information about the three-dimensional spatio-kinematic structure of the nebula, in particular its inner cavity.  
Features are revealed that are not apparent in previously published longslit spectra (Fig.~4 of \citealt{Guerrero:2003a}), which span only the minor and major axis of the nebula. However, there is an error of slit orientation in Figure 4 in Guerrero et al. (2003) for the major axis. They found the cavity on the red side appears to NW and on the blue side to the SE. This is the opposite of what we see. We made a comparison with Sabbadin et al. (1985), and we are completely consistent these authors.

The first-order structure of all the position-velocity images is of a pair of nested closed ellipses, corresponding to the inner and outer shells, roughly consistent with homologous, radial expansion of a spherical nebula. 
From slit \textit{f}, which crosses the central star, we find a systemic velocity of +3~\kms{} heliocentric, together with maximum expansion velocities of \(39.4\)~\kms{} for \nii{} and \(28.5\)~\kms{} for \Ha, consistent with the values found by \citet{Guerrero:2003a}.

In \nii{} (Fig. \ref{fig:pvobsnii}), the ellipses are always hollow due to the negligible abundance of N\(^+\) in the high-ionization core, which corresponds to small angular displacements from the nebular center and small velocity differences from the systemic velocity.
In \Ha{} (Fig.\ref{fig:pvobs}), the ellipses are completely filled for positions close to the minor axis of the nebula (NE and SW quadrants), such as the eastern (upper) portion of slits \textit{d} and \textit{e}, and the western (lower) portion of slits \textit{g}, and \textit{h}. 
For other positions, on the other hand, especially those close to the major axis (NW and SE quadrants), reduced-intensity holes can be seen over certain velocity ranges in the filled ellipses.

These holes clearly correspond to the cavities seen in the direct images (Fig.~\ref{fig:imagenes}). 
When viewed in detail, it is apparent that their arrangement is not consistent with a cylindrically symmetric bipolar structure.  
For instance, the cavities seem blue-shifted in the lower section of slits \textit{d--f}, but also in the upper section of slits \textit{f} and \textit{g}.  
Furthermore, the structure at the top in slit \textit{g} is split into three separate regions or fingers, while in the upper portion of the following three slits, \textit{h--j}, the cavity is red-shifted and also shows hints of splitting. 
In the lower side of the slits \textit{i} and \textit{j} the cavity is on the blue side.
These spatio-kinematic patterns are investigated in more detail in the following section. 

For those cavity structures in \Ha\, that get close to the outer edge of the inner shell, corresponding features can also be seen in the \nii{} spectra.  
Other irregularities in the \nii{} ellipses appear to be associated with the outer shell and do not show any simple correspondence with the cavities in the inner nebula.

\begin{figure*}
\includegraphics[width=\linewidth, trim=0 0 40 0]
{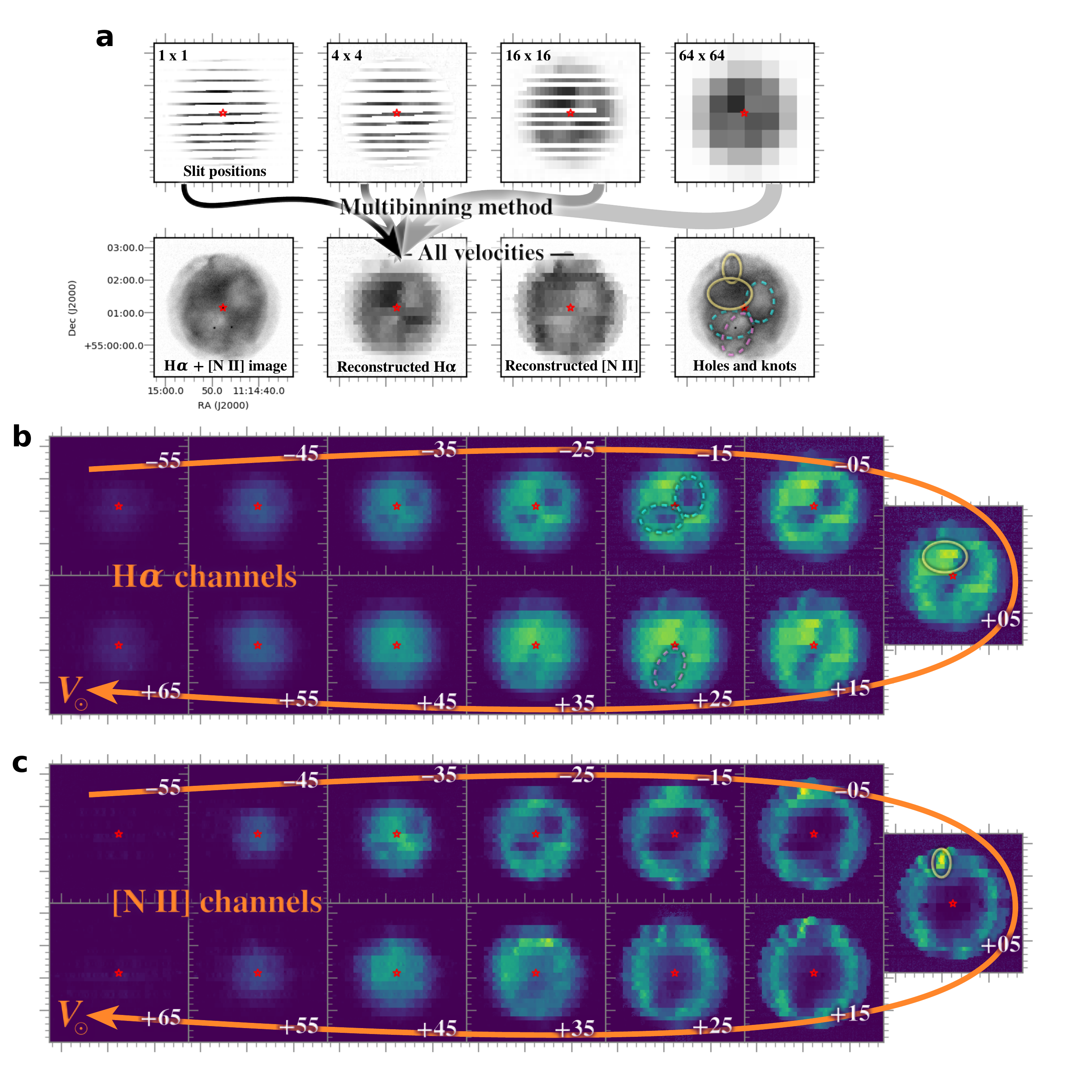}
\caption{
(\textit{a})~Steps in the multi-scale block binning method used for constructing the channel maps from the long-slit spectra (see text for details). 
(\textit{b})~Isovelocity channel maps derived from observed spectra for \Ha{}, shown as false color images, each with 10~\kms{} width. 
(\textit{c})~Same as \textit{b}, but for \nii. }
\label{fig:velocitymaps}
\end{figure*}

\begin{figure*}
  \centering
  \includegraphics[width=\linewidth]{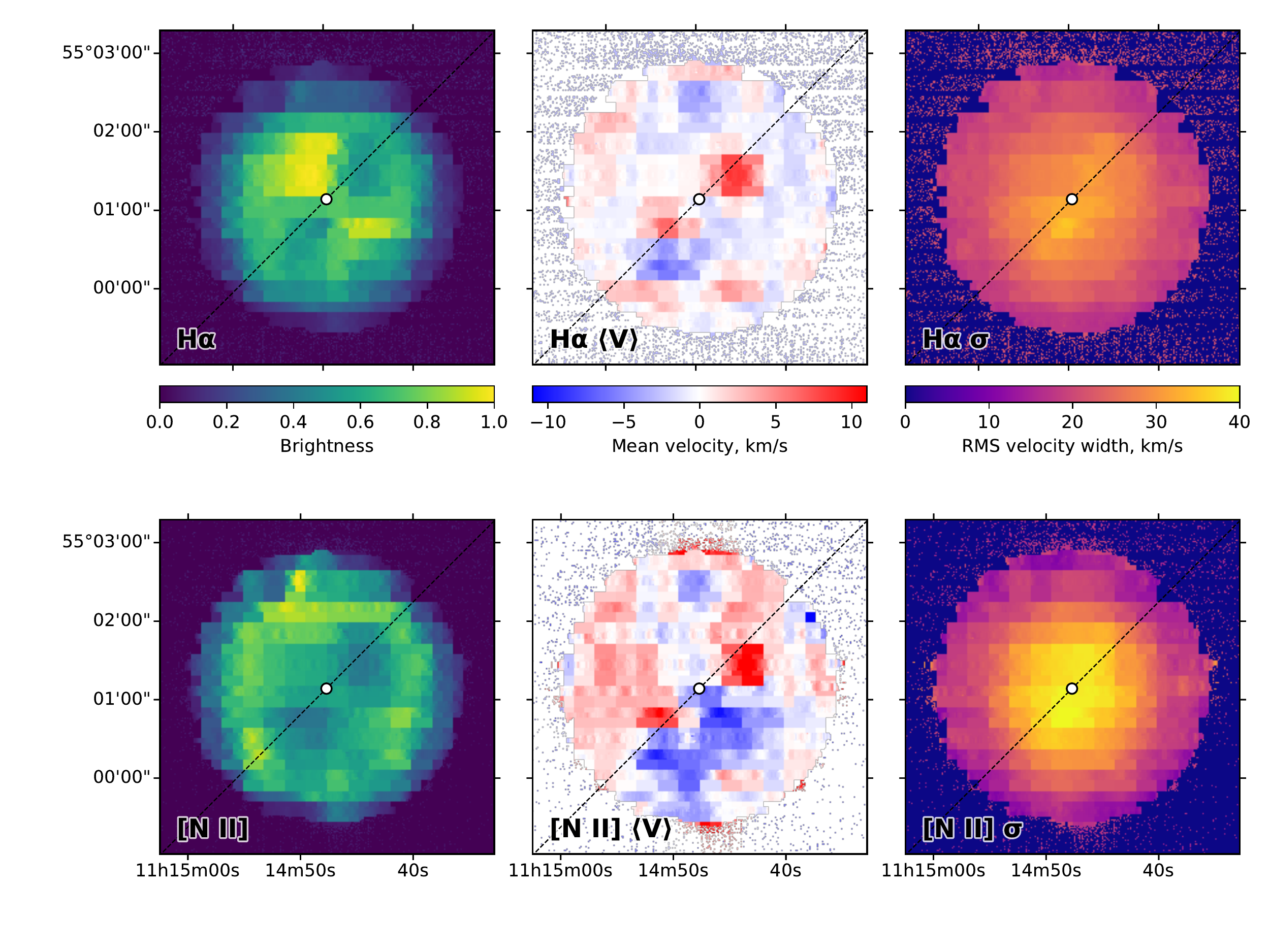}
  \caption{Velocity moment maps of H\(\alpha\) (top row) and [\ion{N}{II}] (bottom row), derived from the isovelocity map reconstruction presented in Fig.~\ref{fig:velocitymaps}. 
  Left to right shows total line intensity, mean velocity, and RMS line width.  
  The white circle shows the position of the central star, while the dashed line shows the approximate major axis of the cavities at \(\mathrm{PA} \approx 315\). }
  \label{fig:owl-moments}
\end{figure*}

\subsection{Isovelocity channel maps}
\label{isovelmaps.sec}

To provide an alternative perspective on our spectroscopic data, we construct iso-velocity channel maps from the long-slit spectra. 
By prioritizing spatial relations on the plane of the sky, channel maps allow a more intuitive understanding than position--velocity images, which helps in interpreting the kinematics and internal structure of the nebula. The model is principally constrained by the P-V arrays and 
direct images. The reconstructed velocity slices are presented merely as a
visualization aid further the block reconstruction does not give 
a uniform spatial resolution across the maps.  The effective resolution 
(in both RA and Dec) at each point is roughly equal to the distance of 
that point from the nearest slit position.  Thus the spatial resolution 
along the slit is not degraded at the slit positions, but it is degraded 
(to about 15\arcsec) at the between-slit positions.


Using the direct image as reference, the long-slit echelle spectroscopic data were relatively flux-calibrated both along and between the slits.  
They were then block-interpolated (see below) to construct position--position channel maps in \Ha\, and \NIIlam{}\AA , as illustrated in Figure~\ref{fig:velocitymaps}.  

The grayscale images in part (\textit{a}) of the figure illustrate the multi-scale block binning method that we use to reconstruct the maps, which is an evolution of techniques described in \citet{Garcia07} and Garc\'ia-D\'iaz et al. (2008a).  
The left panel shows the full-resolution map (\(1 \times 1\)) of the slit brightness profiles on the sky, which shows large gaps in between the slit positions.  
We then repeatedly rebin the map to a \(2 \times\) coarser resolution to produce degraded maps at \(2 \times 2\), \(4 \times 4\), \(8 \times 8\), \(16 \times 16\), \(32 \times 32\), and \(64 \times 64\), where only every second map is illustrated (from left to right).  
This process incrementally fills in the gaps between the slits.  
All the maps are then stacked to produce the final reconstructed image, shown in the second row of Figure~\ref{fig:velocitymaps}\textit{a} for the integrated line profiles, which can be compared with a direct image in the sum of the \Ha{} and \nii{} lines on the left.
The finest resolution map is uppermost in the stack,  but any pixel that contains no data at a given resolution is set to transparent so that the coarser resolution maps will show through from beneath.  

The process was repeated for a sequence of velocity channels  in 10~\kms{} increments from \(-55\) to \(+65\)~\kms{}, with final results shown as color maps in Figures~\ref{fig:velocitymaps}\textit{b} and \textit{c}.
The channel closest to the systemic velocity is shown on the right, while increasingly blueshifted (upper row) and redshifted (lower row) channels are arranged to the left.  
This arrangement is chosen so that front--back symmetries/asymmetries can be easily identified.  
Some notable features in the channel maps are indicated by colored ellipses: dashed lines for holes, solid lines for bright regions.  

Assuming that all gas motions within the nebula are radially away from the central star with a velocity that is linearly proportional to radius, then the spectroscopically measured line-of-sight velocity is linearly proportional to the position along the line of sight. 
This means that that each channel map is effectively a planar slice through the nebula, with blue-shifted channels being in front of the central star and red-shifted ones behind it.
Any departures from such a spherical, homologous expansion would introduce distortions into this simple picture and these are certainly important in younger double-shell nebulae in which the stellar wind is still active, such as the Eskimo \citep{Garcia-Diaz:2012a}.  
However, in an old nebula such as the Owl, such distortions are expected to be much smaller, and we see no evidence for them in our spectra.
A further complication is that thermal, turbulent, and instrumental broadening will introduce a certain amount of leakage between adjacent channels. 
This is especially important for \Ha{}, where the thermal broadening (\(\mathrm{FWHM} \approx 20\)~\kms{}) is larger than the channel widths.

The previously noted bipolar appearance of the inner cavity is clearly visible in the Ha{} channel maps of Figure~\ref{fig:velocitymaps}, as marked by the dashed ellipses in the channels of \(-15\)~\kms{} and \(+25\)~\kms.
However, the nebula does not show the simple red-blue antisymmetry that would be expected from a truly bipolar structure. 
The blue-shifted channels show cavities on both sides of the central star, NW and SE, whereas the red-shifted channels only show a single cavity, on the SE side.
Thus, at least three cavity lobes are required. 
One lies to the NW, pointing towards us, and two lie to the SE: a more easterly one that points towards us, and a more southerly one that points away from us.

\begin{figure}
  \includegraphics[width=\linewidth]{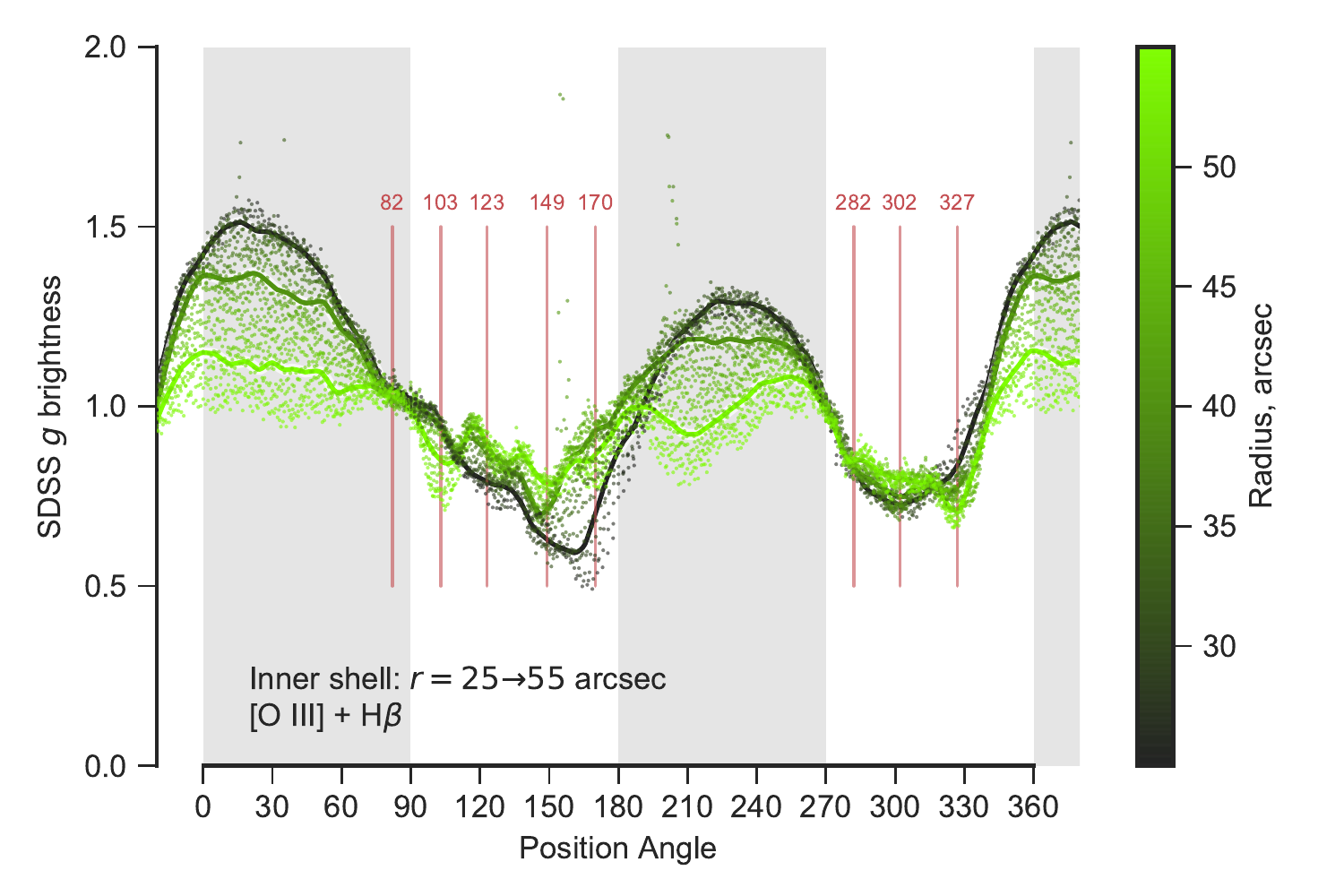}
  \caption{
  Variation of nebular brightness with position angle, measured around the central star. 
  Filled symbols show broad-band brightness (SDSS \(g'\) band, \citealp{Fukugita:1996a}) of individual pixels, with color indicating radius from central star from \(25''\) (black) to \(55''\) (bright green).  
  For this nebula, the continuum is very weak and the spectral band is dominated by the \OIIIlamlam{} and \Hblam{} emission lines.  
  Solid green lines are locally weighted regressions \citep{Cleveland:1988a} in the ranges \(30 \pm 5 ''\),  \(40 \pm 5 ''\),  and  \(50 \pm 5 ''\).
  Vertical red lines indicate the position angles of the fingers used in the SHAPE modeling.
  A white background indicates the PA ranges of the major axis, while a light gray background indicates the PA ranges of the minor axis.
  }
  \label{fig:pa-profiles-green}
\end{figure}

\begin{figure}
 \includegraphics[width=\linewidth]{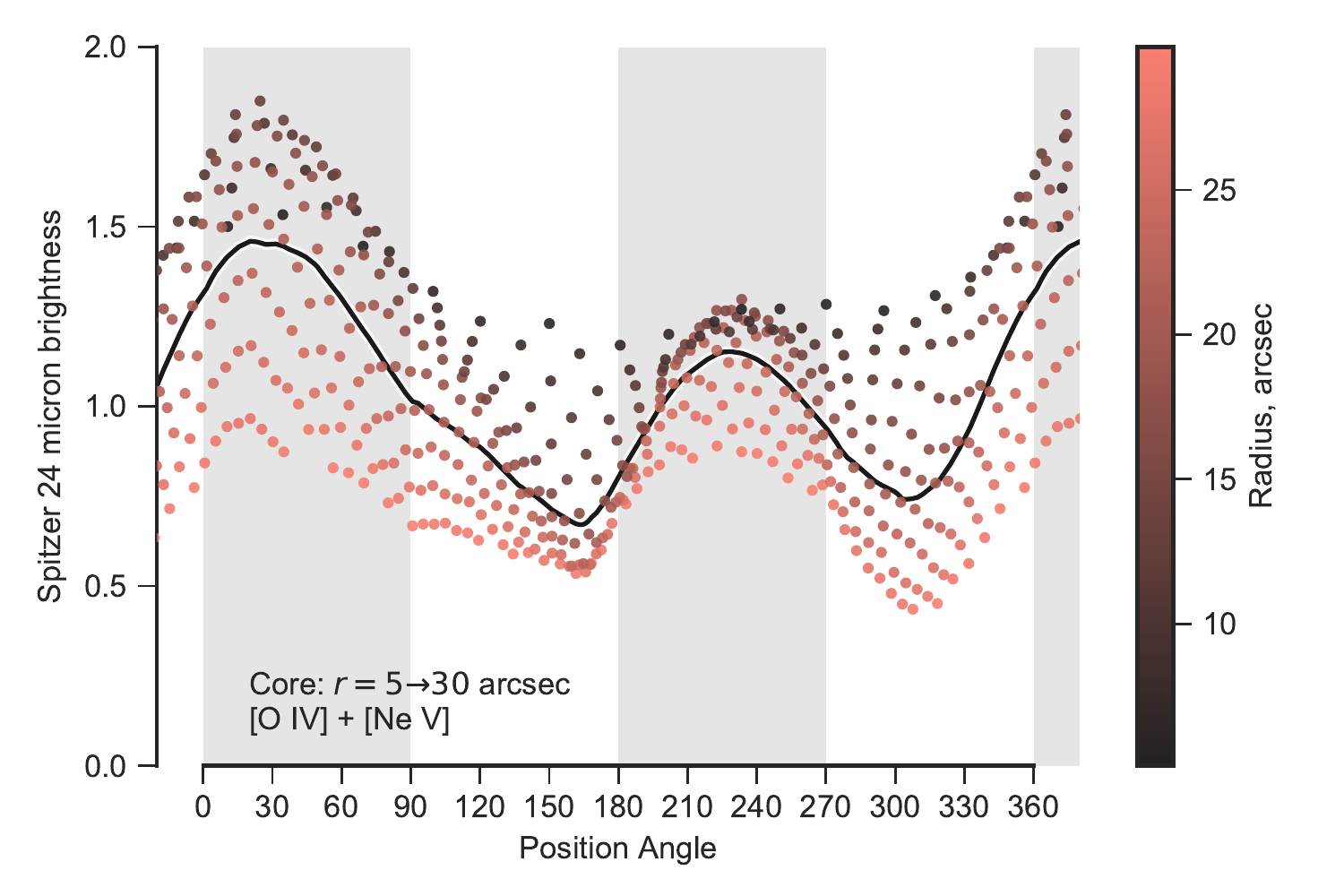}
\caption{As Fig.~\ref{fig:pa-profiles-green} but for the Spitzer MIPS 24~\micron{} band, which is dominated by the [\ion{O}{IV}] \(25.9~\micron\) line.   
Filled symbols show brightness of individual pixels at radii from the central star of \(5''\) (black) to \(30''\) (light red). 
The solid line is a locally weighted regression of all the points.}
  \label{fig:pa-profiles-spitzer}
\end{figure}

\begin{figure}
\includegraphics[width=\linewidth]{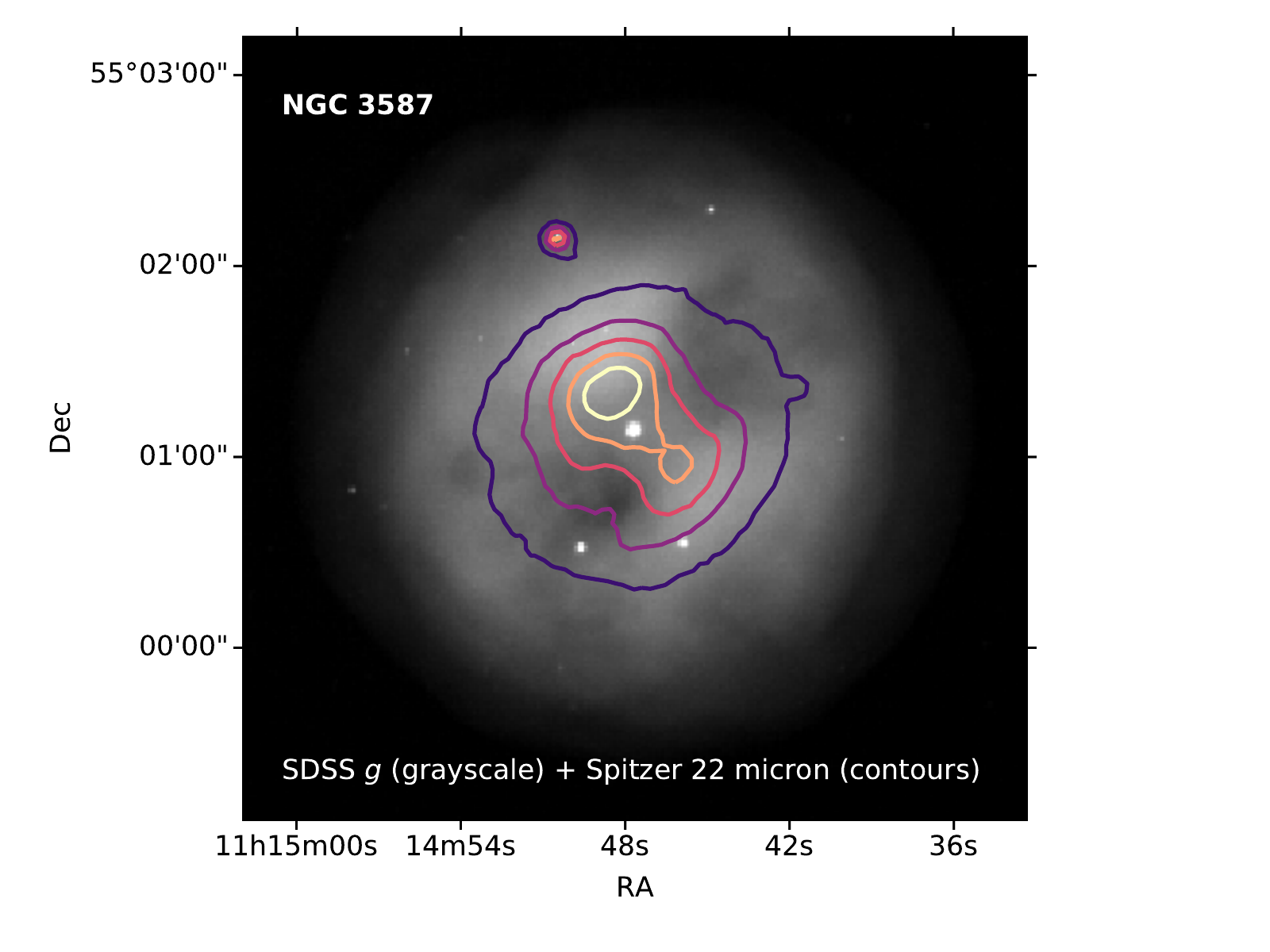}
\caption{High-ionization [\ion{O}{IV}] emission traced by the Spitzer 24~\micron{} band (contours) overlaid on the moderate ionization [\ion{O}{III}] emission traced by the SDSS \(g'\) band (grayscale).}
\label{fig:overlay-owl}
\end{figure}

Similar conclusions can be drawn from the velocity moment maps shown in Figure~\ref{fig:owl-moments}.\footnote{%
Note that a cavity on the blue (near) side produces a \emph{red} shift in the mean velocity, \(\langle V \rangle\), while a cavity on the red (far) side produces a \emph{blue} shift in \(\langle V \rangle\).}
The three lobes described above can be seen as the red and blue patches close to the major axis (dashed line) of the \Ha{} \(\langle V \rangle\) image (top-center panel in Fig.~\ref{fig:owl-moments}). 
Exterior to these inner lobes,  a further pair of blue/red patches can be seen in the mean velocity map, oriented roughly north-south.
These appear to relate to a pair of cavities at \(\mathrm{PA} \approx 0\) and \(\mathrm{PA} \approx 200\) that lie near the outer radius of the inner shell, at \(R \approx  70''\).  
The \nii{} mean velocity map (bottom-center panel in Fig.~\ref{fig:owl-moments}) also shows all the same features, despite the fact that the cavities are not so clearly visible in the \nii{} line profiles of Figure~\ref{fig:pvobsnii}. 
Superimposed on these, the \nii{} \(\langle V \rangle\) map additionally shows a large scale velocity gradient across the minor axis, from red in the NE to blue in the SW. 
By inspecting regions of the \nii{} slit spectra that are well-removed from the major-axis cavities (for instance, upper half of slit e and lower part of slit h in Fig.~\ref{fig:pvobsnii}), it is clear that this is due to a front-back brightness asymmetry in the outer shell, which explains why it is not visible in the \Ha{} line. 

There is even evidence that the three-lobed picture for the inner cavities may itself be an oversimplification.
For instance, some of the \Ha{} slit spectra (especially slit g of Fig.~\ref{fig:pvobs}) seem to show a single lobe splitting in to multiple ``fingers''. 
Similar fingers are apparent on direct images, such as the \oiii{} image shown in the bottom left panel of Figure~\ref{fig:imagenes}. 
At the effective resolution of the channel maps no such  clear split is seen in the north-western cavity. However, the original P-V diagrams reveal more detail, which is consistent with splitting caused by the ``fingers" seen in the images of Figure~\ref{fig:imagenes}.

An alternative way to inspect the structure of the cavities in the Owl is by tracing the nebular brightness variations with position angle around the central star. Figure \ref{fig:pa-profiles-green} shows this variations relative to the SDSS \(g'\) broad-band brightness of individual pixeles in the image.
The graph traces an annulus of radii from \(25''\) (black) to \(55''\) (bright green) from the central star. Vertical lines in the Figure indicate the position angle of the fingers used in the SHAPE model, described in the next section.
We find dips of the order of 10-15\%, which is consistent with the
expectations considering that the thickness of the fingers is roughly
that amount compared to the diameter of the whole nebula. Assuming that
the nebula is of uniform emissivity, the fingers have a moderate angle
to the line of sight, and have low to no emissivity. The measured brightness
dip is somewhat reduced by the seeing, which smooths out density contrast
on scales that are close to the limit of resolution.

Likewise, Figure \ref{fig:pa-profiles-spitzer}  shows, as in Figure \ref{fig:pa-profiles-green}, the variations of brightness with position angle but with relative to the Spitzer MIPS 24~\micron{} band.  Spitzer IRS-LL spectra (Figure~3 of \citealp{Brown:2014a}) show that continuum emission is negligible and the band is dominated by the [\ion{O}{IV}] \(25.9~\micron\) emission line, with a small contribution from [\ion{Ne}{V}] 24.3~\micron{}. The radii of the annulus in this case are from \(5''\) (black) to \(30''\) (light red). Figure \ref{fig:overlay-owl} shows the [\ion{O}{IV}] emission traced by the Spitzer 24~\micron{} band in contours, overlaid on the greyscale SDSS \(g'\) band image. 
Images and spectra attest the reality of the complex cavity structures embedded in the inner shell of the Owl nebula.

\section{Morpho-kinematic modeling with \emph{Shape}}
\label{sec:shape}

The new data presented in this paper show that the central cavity in
NGC~3587 is considerably more complex than the simple bipolar
structure proposed earlier. In order to explore whether this
complexity has any systematic properties that might produce
information about the formation and evolution of the nebula and its
central stellar object, we construct a more detailed
morpho-kinematic 3-D model than has been available before.

We use the morpho-kinematic, polygon-mesh based, 3-D modeling capabilities of
the \emph{Shape} software (\citealt{shapeWo10}; 2017) to model the main
features of NGC~3587, concentrating on the inner dark regions located on the NW and SE sides of the core, forming a bipolar structure described as a dumbbell by \citet{Guerrero:2003a}. However, a simple dumbbell shaped inner cavity does not adequately match the
substructures that we found in our P-V arrays (Figure \ref{fig:pvobs}) and iso-velocity maps (Figure \ref{fig:velocitymaps}), as discussed above. In particular, the southeastern dark region with its split sub structure can
not be matched, nor can the split structures at even smaller scale in
the red and blue-shifted regions. 

The modeling of the complex cavity inside an otherwise filled
H$\alpha$ emission sphere required a somewhat unconventional emission
model compared to previous morpho-kinematic models that used
\emph{Shape}. We therefore describe this procedure in a bit more
detail.

Since we are dealing with a complex cavity, a simple shell model was
unsuitable. Rather, the multi-polar cavity has to be carved out
from a filled sphere. Therefore, the outline of the H$\alpha$ emission
was modeled with a slightly prolate ellipsoidal spheroid that was
filled with uniform emission. Then the multi-polar cavity structure
was built as a polygon-mesh volume. However, instead of giving this volume
a physical emission property, this volume multiplies any emission
inside the volume by some factor smaller than one. It turns out that
the volumes are not completely devoid of emission, therefore the
multiplier is set to 0.3, constant over the full multipolar
structure. This factor is likely to vary considerably over the whole
structure, but trying to model this in more detail would have been too
complex for technical reasons and would not affect the nature of the
result that we obtained. 

\begin{figure*}
\begin{center}
\includegraphics[width=1\textwidth]{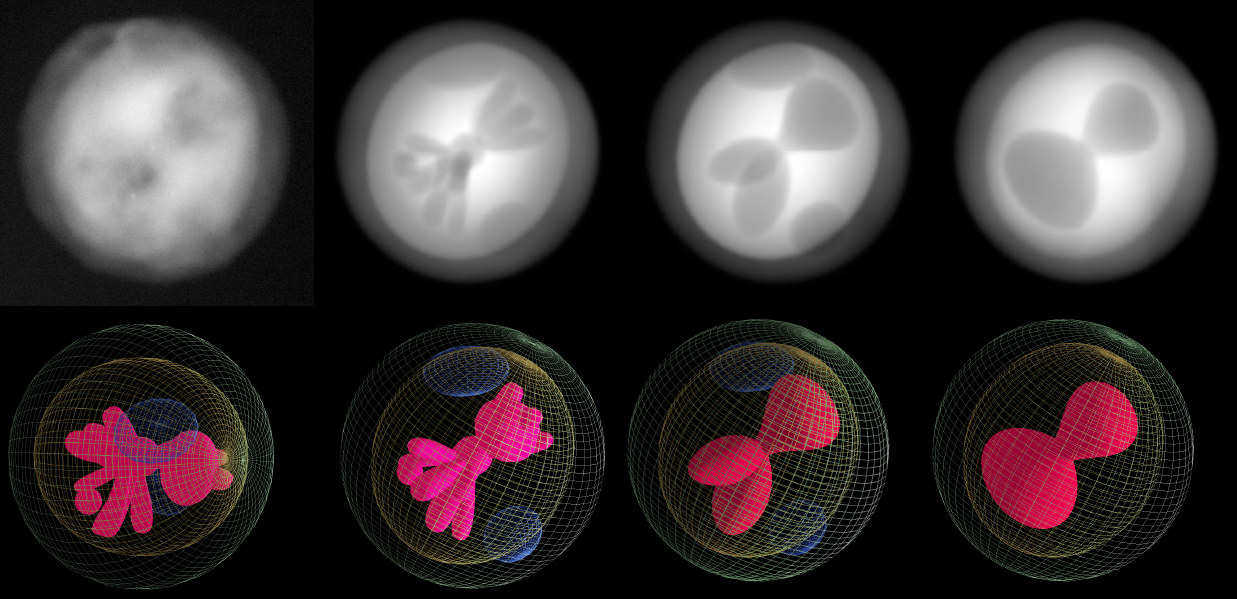}
\caption{A H$\alpha$ image of NGC~3587 (top, left) is compared to three grey-scale synthetic H$\alpha$ image (top, right) derived from the multipolar, tripolar and bipolar models, respectively. Note the matching structure of the multipolar inner cavities. At the bottom the corresponding 3-D multipolar mesh model structure is shown with two different viewing directions {\it first panel:} view from within the plane of the sky with the camera located toward the west of the nebula, {\it second panel:} view from Earth. {\it third panel:} 3-D mesh of the tripolar model image (view from Earth) and {\it Fourth panel:} 3-D mesh of the bipolar model (view from Earth)}
\label{fig:modelo}
\end{center}
\end{figure*}

Figure \ref{fig:modelo} compares our observed \Ha{} image with
the synthetic model image and in Figure \ref{fig:pvobs} the synthetic
P-V diagrams of different models (multipolar, tripolar and bipolar) 
are shown along with the observations (top). Clearly
a bipolar cavity structure is insufficient to explain the detailed
structure of the image and P-V diagram. In particular, note the
change from blue to red-shifted cavities in the southern half from the
P-V diagrams {\it g} to {\it h}, which was impossible to model with a
single connected structure such as a bipolar dumbbell (bottom model), 
even if it was wide ranging from blue to red in terms of Doppler-shift. 
We therefore attempted a model with the south-eastern lobe split into two components, one blue-shifted and the other red-shifted (see the tripolar model in Figures \ref{fig:pvobs} and \ref{fig:modelo}).

\begin{figure*}
\begin{center}
\includegraphics[width=1\textwidth]{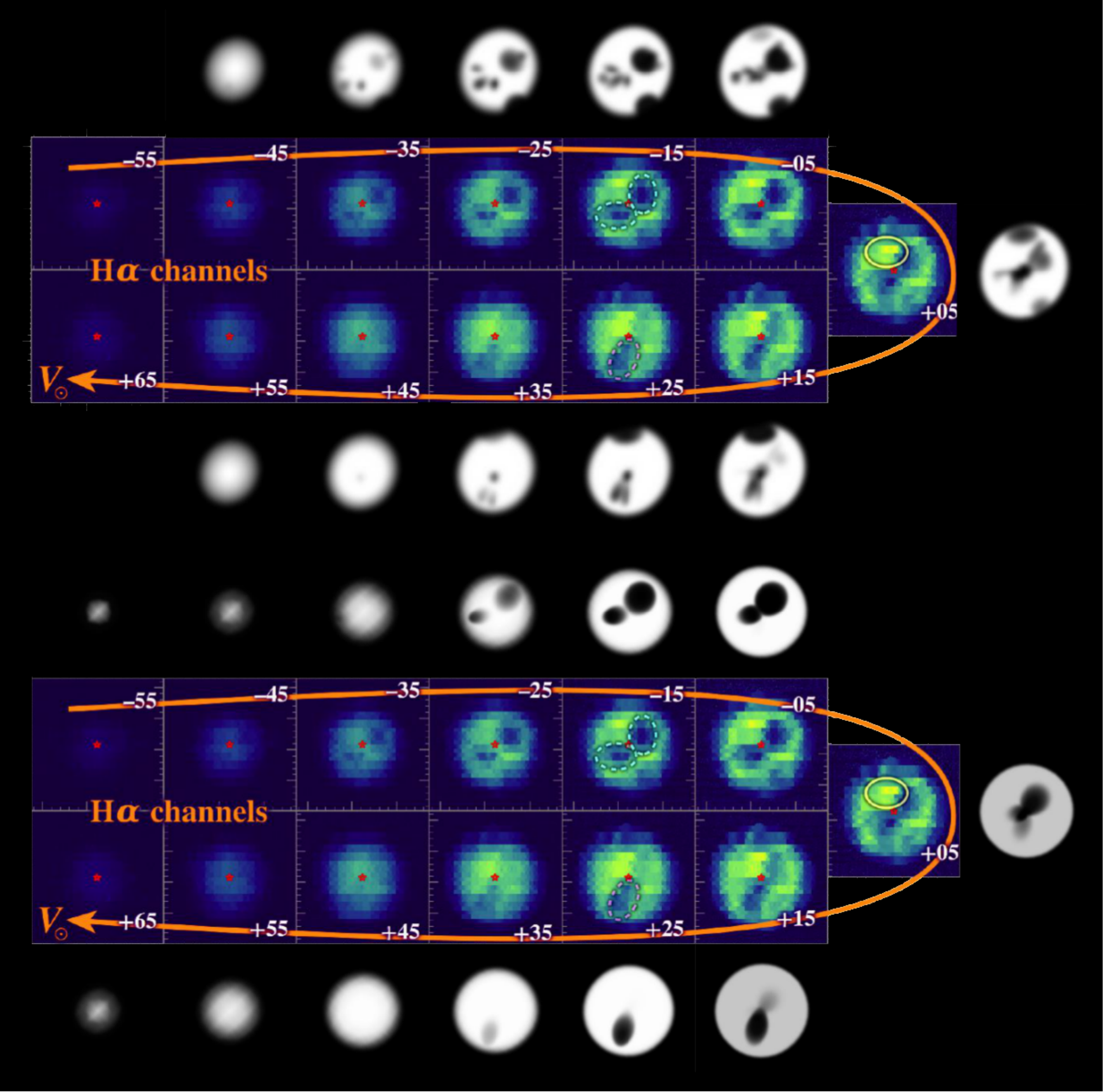}
\caption{Observed isovelocity channel maps shown as false color images reconstructed from the Ha\, long-slit spectra are compared
with synthetic isovelocity channel maps for the multipolar (top) and tripolar (bottom) models.}
\label{fig:isomaps_modelos}
\end{center}
\end{figure*}

This solves the overall structure rather well. Figure \ref{fig:isomaps_modelos} shows the isovelocity channel maps constructed from the observations, as in Figure 5b, these are compared with the synthetic isovelocity channel maps for the multipolar and tripolar models derived from SHAPE. It is clear from this Figure that the tripolar model reproduces the overall structure well. In the bottom part of this figure, the channel maps around $-$15 and $+$15 \kms\, most clearly manifest the separation of the red and blue sections of the south-eastern cavities. In each of these channels only either the blue or red section appears, respectively. 

The remaining significant systematic discrepancies, the finger-like structures, were modeled by splitting the tripolar lobes into smaller components (see the multipolar model images, P-V diagrams and channel maps in Figures \ref{fig:modelo}, \ref{fig:pvobs}, and \ref{fig:isomaps_modelos}), respectively. 
In identifying the fingers, we were guided by publicly available higher contrast and higher resolution imaging (see footnote~\ref{fn:calar-alto} in \S~\ref{sec:imaging}). 
Since there is a danger that image processing steps designed to enhance the resolution might introduce spurious artifacts, we have cross-checked against radially integrated azimuthal profiles of an unprocessed image (see Fig.~\ref{fig:pa-profiles-green}) to ensure that all the features we identify are real.
In the NW group we can identify three different fingers immersed in
a larger bubble-like structure that forms the corresponding lobe in the earlier bipolar interpretation of the cavity structure. 
In the SE, the blue group has at least four fingers, whereas we find three
in the red-shifted group, plus an uncertain smaller fourth. The direction of the fingers used in the model have been marked in Figure \ref{fig:pa-profiles-green}, where they appear as dips in brightness.

In Figure 4 where additional regions of
low emission at the edge of the main nebula have been marked with arrows.
In the more complex (multipolar and tripolar) models these have been
included as simple ellipsoidal regions of reduced emission, which are
indicated as blue colored meshes in Figure 10. They are
likely to be more complex but are not the central subject of this
study and have been included only to make sure they are not part of
the inner multipolar cavity. 

The images and P-V diagrams in Figures 1 and 4 at
first sight do not show any evidence for an inclined elliptical or
bipolar overall structure of the nebula. However, we found that using
a prolate spheroidal shape for the inner shell improves the match
between model and observation. The ratio between the major and minor axes 
of the spheroid is about 1.13 with a position angle of about 60$\degr$ and an inclination of approximately 25$\degr$, with the NW half
pointing towards Earth. This is consistent with the value 1.12
from the extent of the inner shell axis radii found by \citet{Guerrero:2003a}.  The axis of the ellipsoid is roughly aligned with the two
multipolar groups of fingers, within $\pm 20\degr$, that are themselves building an
approximate overall bipolar structure.

\section{Discussion}
\label{sec:discussion}
\subsection{The Owl in the context of PN populations and evolution}
\label{sec:discuss-context}
An approximate kinematic age, \(t = R / \dot{R}\), can be derived for the Owl Nebula from the shell radius, \(R\), and expansion velocity, \(\dot{R}\).  
Using the recommended prescription for old nebulae, \(\dot{R} \approx 1.3\, V_{\mathrm{[\ION{N}{2}}]}\) \citep{Jacob:2013a}, yields \(t = 8400\)~yr.  
Taken at face value (but see \citealp{Schonberner:2014a}) and, combined with the central star evolutionary tracks of \citet{Miller-Bertolami:2016a}, this age requires a remnant stellar mass in the range \(0.58\)--\(0.62~\mathrm{M_\odot}\) (initial mass \(1.5\)--\(2.5~\mathrm{M_\odot}\)) and a metallicity closer to \(Z = 0.01\) than \(Z = 0.02\).    
The ionized nebular mass of the different nebular components can be found from the densities and sizes derived in \S~\ref{sec:introduction}, yielding \((0.36 \pm 0.18)~\mathrm{M_\odot}\) for the inner shell, \((0.41 \pm 0.20)~\mathrm{M_\odot}\) for the outer shell, and   \((0.22 \pm 0.11)~\mathrm{M_\odot}\) for the halo.   Summing these three components, together with the present-day stellar mass yields a total of \((1.6 \pm 0.3)~\mathrm{M_\odot}\), which is consistent with the evolutionary tracks.  (The initial mass could be slightly higher if there were a significant amount of undetected neutral gas in the halo.)

This implies that the Owl progenitor comes from the intermediate mass population of the thin Galactic disk, with a total evolutionary age of \(\sim 5\)~Gyr.
The same conclusion follows from the observed nebular abundances (\(\mathrm{He} / \mathrm{H} = 0.095\), \(\log_{10} \mathrm{N} / \mathrm{O} = -0.60\)) and height above the Galactic disk (\(z = 0.5\)~kpc), which lead to a classification as Peimbert Type~IIb \citep{Quireza:2007a, Peimbert:1978a}. 

\subsection{The Owl as prototype of the family of strigiform nebulae}
\label{sec:discuss-context}

Bipolar and multipolar structures are common in planetary and pre-planetary nebulae, but in most cases are of a very different nature to those seen in the Owl. 
Evolved nebulae that are classified as bipolar tend to be nitrogen- and helium-rich (Peimbert Type~I), indicating a relatively high mass (\(> 2.4~\mathrm{M_\odot}\)) and young (\(< 2\)~Gyr) progenitor \citep{Phillips:2003a, Maciel:2011a}.  When observed in detail, many supposedly bipolar nebulae turn out to be quadripolar or multipolar \citep{Manchado:1996a, Hsia:2014a}, especially those that are small and young \citep{Sahai:2011a}.
Many show jet-like lobes that are indicative of complex dynamic interactions involving collimated fast outflows, shocks and/or ionization/dissociation effects \citep{Garcia-Segura:2010a}, which  may be related to the presence of a putative binary central star \citep{Sahai:2005a}.
 
 The Owl nebula represents a very different phenomenon that has been largely ignored in the models. The Owl is a large,  evolved, filled nebula, surrounding a low-luminosity star, which shows no evidence for a present-day stellar wind. 
 Yet, it exhibits at this late stage in its evolution a complex asymmetric inner structure, where multiple cavities have been carved and whose overall shape form the ``eyes'' of the Owl nebula. 
Interestingly, the Owl nebula is not the only PN with these characteristics. 
Its twin southern counterpart is found in K 1-22 (aka ESO 378-1), while Abell 33 and Abell 50 are other PNe that share many of the particular characteristics mentioned above, see Figure 12. That Figure also shows single position, long-slit echelle spectra for these nebulae, drawn from the SPM Kinematic Catalogue of Galactic Planetary Nebulae (\citealt{LOpez:2012}). The shapes of the spectral lines and the nebular morphology are very similar to NGC 3587.  
We propose that these objects constitute the new class of \emph{strigiform} nebulae,\footnote{%
The name ``strigiform'' is apt because of its zoological meaning: ``of, relating to, or belonging to the Strigiformes, an order of birds comprising the owls'' \url{https://en.wiktionary.org/wiki/strigiform}.
} 
and we intend to carry out a comprehensive spectral mapping of these objects in the near future to compare with the Owl data.   

\begin{figure*}
\begin{center}
\includegraphics[width=0.7\textwidth]{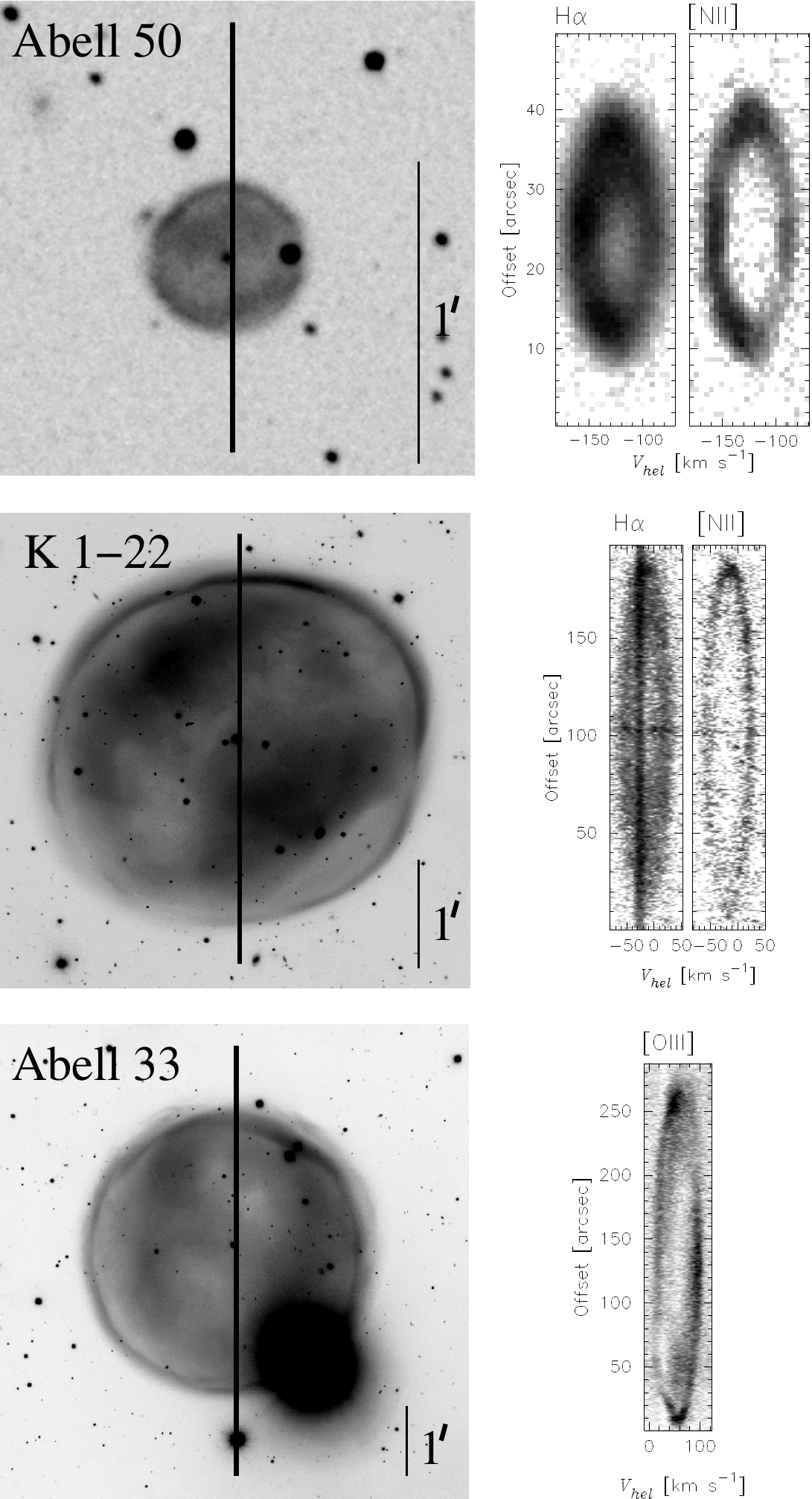}
\caption{Other PNe in the strigiform class: Abell 50, K 1-22 and Abell 33.  For each nebula, we show a direct image in H$\alpha$ + \nii{}, together with single-position long-slit echelle spectra (L\'opez, et al. 2012) }
\label{fig:family}
\end{center}
\end{figure*}

\begin{figure*}
\setlength{\tabcolsep}{0pt}
\begin{tabular}{ll}
(a) & (b) \\
\includegraphics[width=0.5\linewidth]{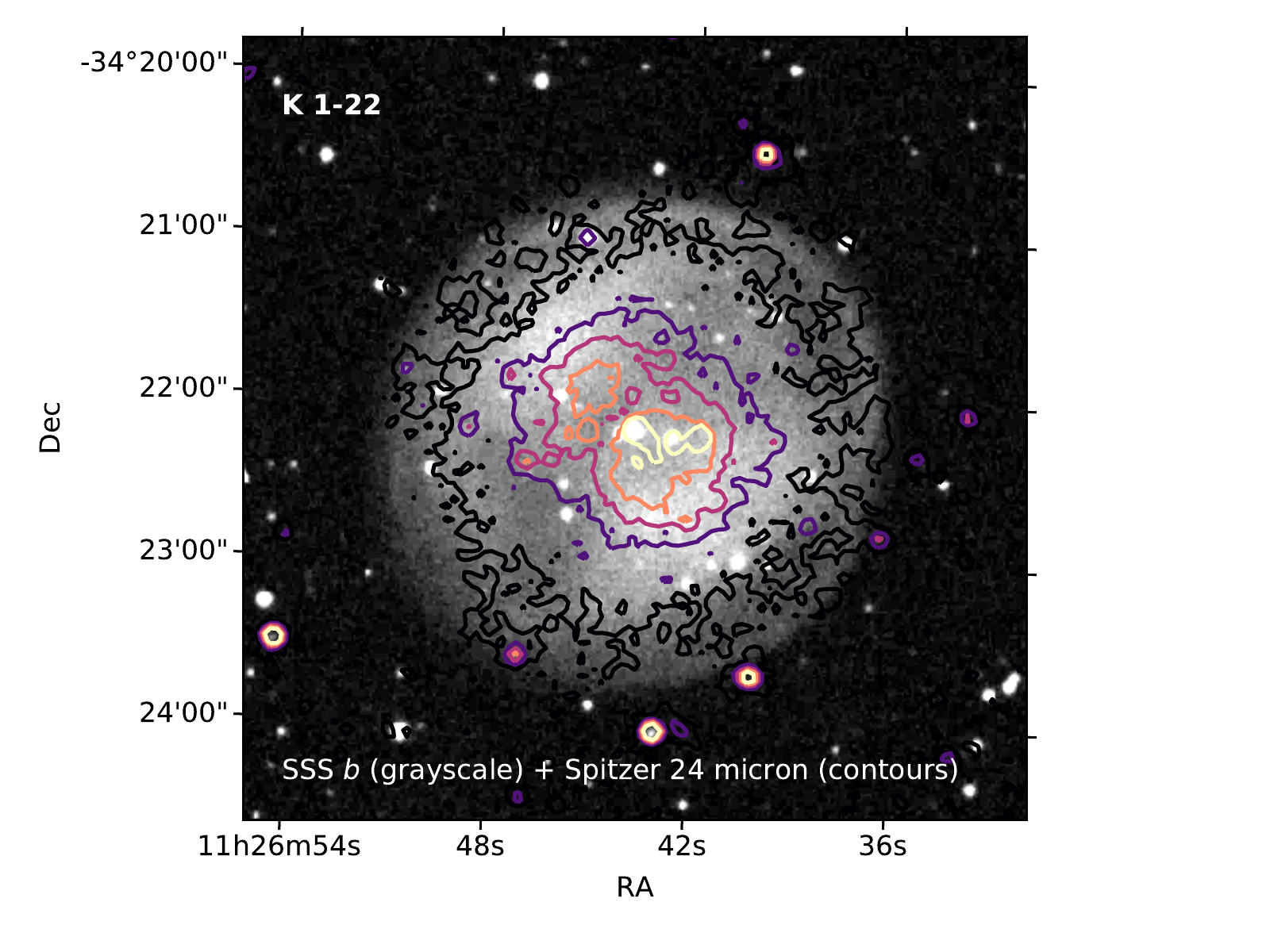} &
\includegraphics[width=0.5\linewidth]{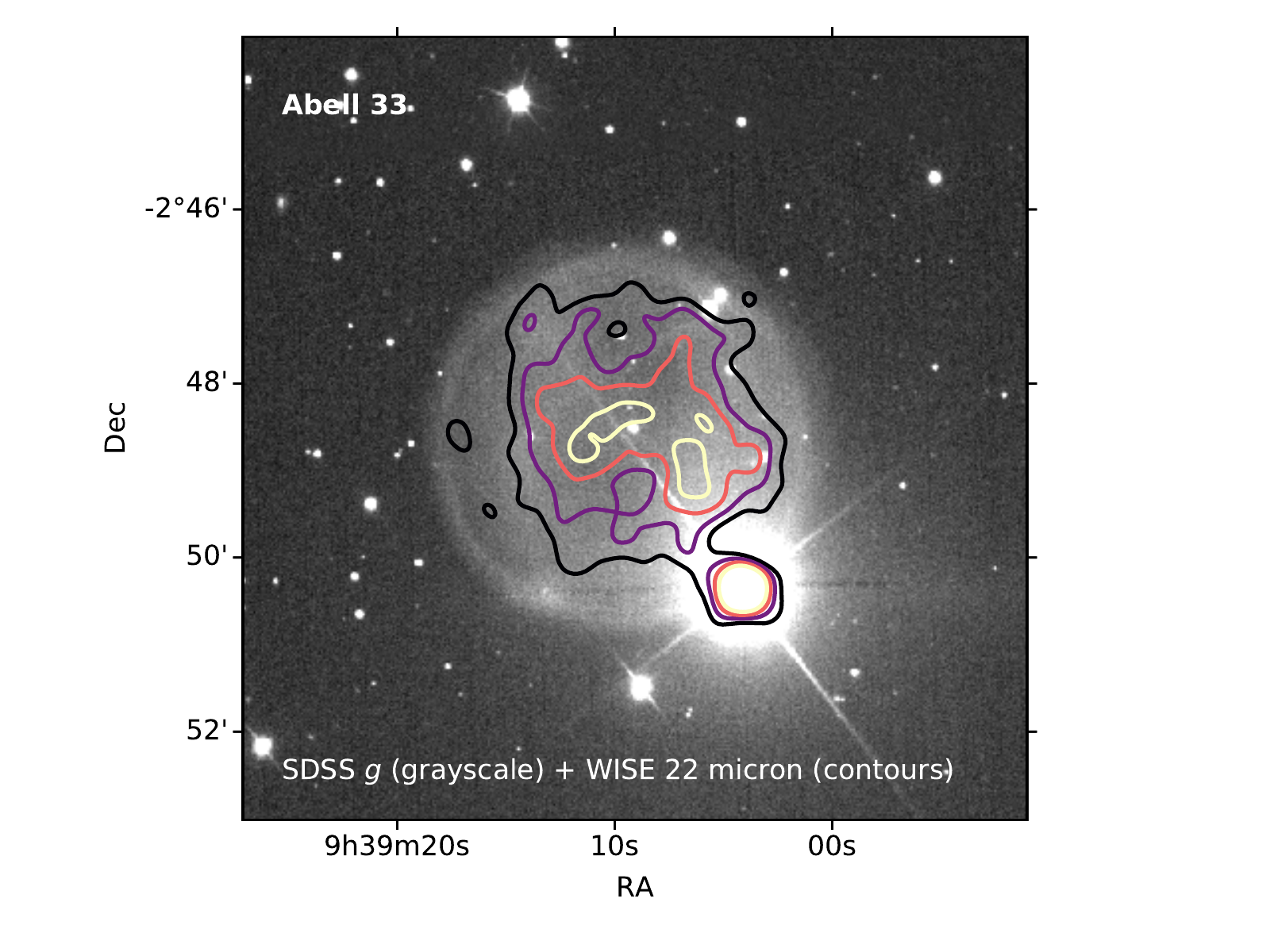}
\end{tabular}
\caption{Comparison of mid-infrared emission and optical emission for strigiform nebulae. (a) K~1-22: contours show Spitzer \(24~\micron\) emission, grayscale shows SuperCOSMOS Sky Surveys blue broadband emission. (b) Abell~33: contours show WISE band 4 (\(22~\micron\)) emission, grayscale shows SDSS green broadband emission.}
\label{fig:strigiform-24micron}
\end{figure*}

These PNe share additional similarities with the Owl. The central stars are all close to \(L = 100~\mathrm{L}_\odot\) and  \(T_\mathrm{eff} = 100\)~kK \citep{Frew:2008a} and their global expansion velocities are in the order of \(V_{\mathrm{exp}} = 30\)--\(40 \mathrm{\ km\ s^{-1}}\). They belong to the group of PNe that \citet{Pereyra13} call HE or highly evolved  PNe, (see their figure 8). At this very late stage of evolution the stellar wind is presumed to be very weak or have ceased. \citet{Guerrero13} show that for NGC 3587 the latter applies. 
Due to the rapid decrease in the luminosity of the central star after H-burning ceased, the nebulae are likely to have passed through a recombination phase in the past \citep{Tylenda:1986a}, but now that the stellar cooling timescale is longer than the nebular expansion timescale, the main nebular shells will have entered a re-ionization phase, driven by the declining density, although the outer halos may still be recombining.

NGC 3587, Abell 33, and K 1-22  are within a heliocentric distance of 1 $\pm 0.5$  kpc, whereas Abell 50 is more distant, with an estimated distance of 5 kpc (\citealt{Stanghellini:2010}). 
K 1-22 and Abell 33 have very wide companions with orbits in the order of 500 and 2000 AU respectively (\citealt{Ciardullo99}). 
For NGC 3587  there are indications of a binary companion from I-band excess observations \citep{Douchin15} but no additional data for the orbit seems to be available.
For Abell 50,  \citet{DeMarco15} could not find conclusive evidence of variability from Kepler data. 
For these very long-period binaries little influence on the shape of the nebula is expected, although there might be some contribution to turbulence in the inner nebula, as indicated by the width of their H$\alpha$ line profiles that is around 20 km s$^{-1}$ in all cases.  

All of the strigiform nebulae show double shells, but their morphology is very different from the narrow inner rim plus broad outer shell, which is seen in younger double-shell nebulae \citep{Stanghellini:1995a}.  
Instead, it is the inner shell that is broad and diffuse, while the outer shell is narrow and more structured.  In the case of the Owl, the ionized masses of the two shells are approximately equal.  
A theoretical perspective on structure and kinematics of this evolutionary stage is given in \S~6 of \citet{Schonberner:2014a}.

The strigiforms are also very homogeneous in their mid-infrared \(24~\micron\)  emission characteristics, being classified as Group~3, or centrally concentrated, according to the taxonomy of \citet{Chu:2009a}.  
In this group of objects, dust continuum emission is very weak or absent, and the Spitzer and WISE bands are both dominated by high-ionization [\ion{O}{iv}] line emission \citep{Brown:2014a}.  
Figure~\ref{fig:strigiform-24micron} compares maps of the [\ion{O}{iv}] mid-infrared line (contours) with optical emission (grayscale) for two of the strigiforms.\footnote{%
Abell~50 is not included because its small angular size means that its mid-infrared emission is unresolved. }
In both cases, the holes in the optical emission coincide with a reduction in brightness in the infrared, which means that variable dust extinction can be conclusively ruled out as an explanation for the apparent cavities.

All these common elements suggest the existence of a cohesive group of PNe, with common evolutionary paths that have not been properly investigated and warrant further study.  In addition to the Owl and the three nebulae shown in Figure~\ref{fig:family}, other nebulae that are candidates for inclusion in the strigiform family include NGC~6894, K 1-20 and IC 1454, which bear a strong morphological resemblance to the class, but for which we possess no slit spectra.

\subsection{Nature of the cavities}
\label{sec:discuss-cavities}

The inner multipolar cavities that we find in the Owl and the other strigiform nebulae are clearly deficient in photoionized gas with the same density and temperature (\(100~\mathrm{cm^{-3}}\) and \(10^4~\mathrm{K}\)) that is typical of the rest of the nebula. 
The question then arises whether the corresponding shortfall in thermal pressure is made up for by other means, or if the cavities are unsupported structures that are currently undergoing dynamical collapse. 
We find no kinematic evidence in the line profiles for such a collapse, but the low velocities involved (\(\approx 10~\mathrm{km\ s^{-1}}\)) would not be easily detectable. 
However, the size of the smallest scale structures we see in the Owl's fingers is roughly 10\% of the radius of the inner shell, which is expanding at 2--3 times the speed of sound. 
This means that the sound crossing time of these structures is no more than one-quarter of the expansion age of the nebula, so even if the cavities are not actively supported in the present day, they must have been so-supported in the recent past.
The missing pressure in the cavities could conceivably be provided in various forms: ram pressure of active winds or outflows from the star, thermal pressure of a hot bubble, or magnetic pressure.\footnote{%
Other possibilities can be immediately ruled out: direct radiation pressure (central star is insufficiently luminous) and non-thermal particles (no evidence for synchrotron emission).
}
The most plausible origin of all these is the fast stellar wind from the central star. 

There are three principal lines of evidence for the importance of the fast stellar wind in the evolution of planetary nebulae:  kinematic and morphological signatures in the nebular shells, diffuse x-ray emission from the central nebula, and ultraviolet P~Cygni profiles in the spectrum of the central star.  
All three are lacking in the case of the Owl and other strigiform nebulae, indicating that wind activity has ceased or declined to a negligible level.
Many young and middle-aged double shell nebulae show spectroscopic evidence for a complex internal kinematics driven by the fast stellar wind \citep{Guerrero:1998a, Garcia-Diaz:2012a},  but no such features are seen in the strigiform nebulae, where the line profiles (Fig.~\ref{fig:pvobs} and \ref{fig:family}) are consistent with a uniform homologous expansion.  
Similarly,  diffuse x-ray emission, which is evidence for a hot shocked wind bubble is only found \citep{Freeman:2014a} in compact nebulae with ionized densities \(> 1000~\mathrm{cm^{-3}}\), an order of magnitude higher than is seen in the Owl.  
P~Cygni profiles of high-ionization far-ultraviolet absorption lines   \citep{Guerrero13} are direct diagnostics of the wind mass-loss rate, \(\dot{M}\),  and velocity, \(V_\infty\), but these are not detected in the Owl (the other strigiform nebulae have not been observed).
Most detected P~Cygni profiles are from low-gravity, high luminosity central stars \citep{Pauldrach:1988a, Hoffmann:2016a},  where the wind is strong (\(\dot{M} =  10^{-8} \text{--} 10^{-6}~\mathrm{M_\odot\ yr^{-1}}\)).  
The steep dependence of \(\dot{M}\) on luminosity for radiatively driven winds \citep{Lamers:1999b} makes it much harder to detect in more evolved stars.
The most evolved CSPN where mass loss estimates have been made are \(\dot{M} > 7 \times 10^{-11}~\mathrm{M_\odot\ yr^{-1}}\) for LoTR~5 \citep{Modigliani:1993a} and \(4 \times 10^{-12} < \dot{M} < 10^{-10}~\mathrm{M_\odot\ yr^{-1}}\) for NGC~1360 \citep{Herald:2011a}, but these are still about 10 times more luminous and lower gravity than the strigiform central stars.
Extrapolating the wind--luminosity relation would imply \(10^{-13} < \dot{M} < 10^{-11}~\mathrm{M_\odot\ yr^{-1}}\) in the strigiforms, but the wind may be even weaker if decoupling  of hydrogen from the metals occurs, as is predicted for sufficiently high gravity \citep{Unglaub:2007a}. 

It thus seems likely that the pressurization of the strigiform cavities is a relic of an earlier high-luminosity phase of the CSPN evolution, when the stellar wind was much stronger.  
The dynamical evolution of nebulae in this phase, after the termination of the strong fast stellar wind, is explored via numerical simulations in \citet{Garcia-Segura:2006a},  where it is found that fragmentation of the earlier wind-driven shell leads to the formation of a system of evaporating dense knots, which are similar to those seen in the Helix and other evolved nebulae \citep{ODell:2002a}. 
However, dense knots and molecular hydrogen emission are found preferentially in nebulae with relatively high initial stellar masses, which tend to show enriched He and N abundances and bipolar nebular morphologies. 
No such knots are seen in the strigiform nebulae, which likely come from lower mass progenitors (see \S~\ref{sec:discuss-context}). 

Models that describe multipolar structures in young planetary and pre-planetary  nebulae mainly go along one of two lines. Either episodic collimated outflows from a single or binary central system (e.g. \citealt{LOpez:1993}) or the expansion of an isotropic fast wind through a dense, inhomogeneous circumstellar post-asymptotic giant branch wind (\citealt{SKE13}) 
However, it is clear that none of these models, or their variants apply here.
Furthermore, the cavities forming the lobes show no evidence of significant edge-brightening that would point towards them being the result of expanding shocks that had swept up the gas, as would be expected if an active stellar wind or a passing collimated outflow had been the cause of their formation. 
 
On the other hand, the multipolar nature of the strigiform cavities is also hard to explain by an isotropic collapse of thermal pressure in the stellar wind bubble. 
In the case of the Owl, our 3D modeling (\S~\ref{sec:shape}) implies that the cavity radius varies by a factor of more than 10 between different directions from the star. 
The situation is not so clear in the case of the other strigiforms, but at least for K~1-22, the appearance of the [\ion{O}{iv}] emission (Fig.~\ref{fig:strigiform-24micron}) implies a very similar morphology. 
Magnetic fields could potentially account for this anisotropy in one of two ways:
directly in the case where magnetic pressure and tension dominate the pressure of the bubble (e.g., \citealp{Chevalier:1994a}), or indirectly by suppressing thermal conduction perpendicular to the field lines \citep{Balsara:2008b}, but the magnetic field topology would need to be more complicated than a simple dipole or split monopole.

Alternatively, the asymmetries in the cavities may themselves be relics of asymmetries in the wind-driven shell during prior evolutionary stages. 
The class of middle-aged elliptical planetary nebulae with ansae, such as NGC~2392, 6543, and 7009 \citep{Garcia-Diaz:2012a} may be a guide in this respect.  
These show a highly elongated, fast-expanding, wind-driven rim, embedded in a more spherical, slowly expanding outer shell. 
The subsequent evolution of such structures after the termination of the fast wind, might potentially explain the strigiform cavities, although it remains to be seen whether the multiple fingers seen in the Owl could arise naturally from such a mechanism. 

Another possible precursor to the strigiform nebulae is provided by NGC~1360 (Garc\'ia-D\'iaz et al. 2008b, Miszalski et la. 2018a).  
This resembles the strigiform class in that is essentially a filled ionized volume, containing a low-contrast, highly asymmetric cavity, which is also apparent in the [\ion{O}{IV}] mid-infrared line \citep{Chu:2009a}.  
It differs from the strigiforms in containing a higher luminosity central star (\(4000~\mathrm{L_\odot}\), \citealp{Traulsen:2005a}) and in being significantly more prolate.  
In addition, NGC~1360 shows small-scale structures outside the main nebular body: fast-moving knots on the prolate axis (Garc\'ia-D\'iaz et al. 2008b) and low-ionization knots around the equator \citep{Miszalski:2018a}, which may be related to its putative post-common-envelope binary nature.   
The shape of the cavity in this object is reminiscent of what would be expected from a precessing jet, suggesting a relationship to the fast axial knots. 
If the strigiform cavities were generated similarly, then this might indicate a connection with binarity in those objects too, although the exact mechanism remains unclear.
In the Owl, there is no evidence for emission knots outside the nebular body, although the outer [\ion{N}{II}]  emission shell does seem to bear an imprint of the inner-shell cavities (see Fig.~\ref{fig:pvobs}), in that the outer [\ion{N}{II}] brightness shows slight dips at the position angles of the fingers.
However, this may simply be an illumination effect, due to the reduced density in the cavity allowing a greater ionizing flux to reach the outer shell, hence reducing the \(\mathrm{N}^+\) abundance there.

\section{Conclusions}
\label{sec:conclusions}

We propose that NGC 3587 be recognised as the prototype of a new class of \textit{strigiform} nebulae, which are highly evolved PNe that show complex, multipolar cavities in the inner regions of an otherwise nearly spherical, filled  nebula. 
We find three further clear members of the family (K~1-22, Abell~33, and Abell~50), where the morphology of mid-infrared [\ion{O}{IV}] emission shows that the cavities are true holes, and are not caused by uneven dust extinction. 
The cavities show no internal rims that might indicate active shock activity and the
central stars show evidence for a very weak or absent present-day stellar wind,  which suggests that the cavities represent relics of an earlier strong-wind, high-luminosity phase of evolution, but their strongly asymmetric nature presents a challenge to theoretical models.
Additional detailed observations of these objects are planned for the near future, as well as follow-up studies on other potential members or precursors of the strigiform class, such as NGC~6894, K~1-20, IC~1454, and NGC~1360. 
We hope that this work will stimulate future nebular modeling of very late stages of planetary nebula evolution.

\section*{Acknowledgements}
We thank the daytime and night support staff at the OAN-SPM for
facilitating and helping obtain our observations, particularly Gustavo
Melgoza, Felipe Montalvo, Salvador Monroy and Francisco Guillen, who
were the telescope operators during our observing runs.
The authors acknowledge support from UNAM grants DGAPA PAPIIT IN110217,
101014, 104017, 108416.

\bsp

\label{lastpage}

\end{document}